\documentclass{article}
\usepackage{graphicx}
\usepackage{subcaption}
\usepackage[T1]{fontenc}
\usepackage[utf8]{inputenc}
\usepackage[english]{babel}
\usepackage{amsmath}
\usepackage{mathrsfs}

\usepackage[left=2cm,right=2cm,top=2cm,bottom=2cm,bindingoffset=0cm]{geometry}

\usepackage[natbib=true, backend = bibtex, style=numeric-comp, sorting=none]{biblatex}
\title{Quantum and classical symmetries}
\author{A.Anokhina\footnote{NRC ``Kurchatov Institute'', Moscow; Institute for Information Transmission Problems; anokhina@itep.ru}, M.Satleikin\footnote{M. V. Lomonosov Moscow State University, Moscow; satleikin.mi21@physics.msu.ru}, A.Sedova\footnote{M. V. Lomonosov Moscow State University, Moscow; anastasiia.sedova@math.msu.ru}, A.Shukakidze\footnote{Tbilisi State University, Tbilisi; alexander.shukakidze534@ens.tsu.ge}.}
\date{}

\begin{document} 

 \numberwithin{equation}{section}
	\maketitle

\vspace{-5cm}
\begin{flushright}
ITEP-TH-17/24\\

IITP-TH-14/24
\end{flushright}
\vspace{4cm}

	\begin{abstract}
		We suggest a somewhat non-standard view on a set of curious, paradoxical from the standpoint of simple classical physics and everyday experience phenomena. There are the quantisation (discrete set of values) of the observables (e.g., energy, momentum, angular momentum); forbidden simultaneous measurements of the observables in the most cases (e.g., of a coordinate and momentum, of angular momentum projections on difference axis); counter-intuitive relations on the simultaneously measurable quantities (e.g., the famous expression for the square momentum $l(l+1)$ with the maximal projection $l$). 
		These and other paradoxes are traditionally related to ``purely quantum'' phenomenon, i.e., having no analogue in the ``classical world'' ones. However, there are deep analogies between classical and ``quantum'' worlds, as soon as the quantum technique is applied to the classical phenomenon. We follow these analogies with the examples of relatively simple and well known models of classical physics, such as a simplified model of light transition through the media, a system of electric charges close to each other and far from the observer; the specific of motion in the Coulomb/Newtonian field.
		
		This text can be considered as a mini-course addressed to higher school and undergraduate students who are interested in basics of quantum mechanics, but are not yet ready for systematic study of standard courses. The text may be also useful to those who supervise such students.
	\end{abstract}

\section{Introduction}
\subsection{Physics of simple and complicated systems. (In)determinism}
Relatively simple physics systems, such as a material point in external field, an absolutely solid body with a fixed point, a homogeneous absolutely elastic media, a set of fixed electric charges or magnetic points, a constant or sinusoidal current circuit with ideal resistors, inductors, and capacitors, an ideal gas, a quasi-equilibrium phase transition liquid--solid or liquid-gas, a system of thin lenses, prisms and polarisers, ect., are deterministic. I.e., it is \textit{principally possible} to measure all physical quantities associated with the system at each moment, and it is \textit{principally possible} to predict (on the grounds of measurement data and physical laws) to predict exact evolution of each of the quantities in time.

However, just a bit	more complicated systems, such as an asymmetric top on a rough surface with a free point or even contact patch, a paramagnetic--ferromagnetic phase transition, a non-equilibrium heat process, diffraction and other physical, in particular non linear optics, a radiation and a propagation of waves (e.g., mechanical or electromagnetic), already cant be considered as purely deterministic. I.e., it is \textit{principally impossible} (or extremely complicated and meaningless) either to measure all physical quantities associated with the system, or to predict exact evolution of each of the quantities in time. Physical description on such systems contains \textit{irreducible uncertainty}. I.e., it does not tend to describe the system in all imaginable details. Opposite to that, the description accepts as essential property of the system that out knowledge about the system is restricted. 
	
Of course, all really complicated systems, such as atomic nuclei and subatomic particles, solid and liquid media, turbulent flows, a weather and a climate, models of vital systems, ect. are highly indeterminate. Sources of the indeterminism vary from a big number of degrees of freedom and chaotic evolution of the system to irreducible impact of the observer on the system. The art of physics of indeterminate systems starts from searching for proper quantities that are physically measurable and predictable in their evolution and that at the same time represent essentially properties of the system \cite{Hack}. And all these peculiarities are present in quantum mechanics, already in its simplest variant, which includes one particle, stationary system and finite spectra. In this sense quantum mechanics is not a marginal model of electron, which ``one can not understand, but can only use'', but a huge step towards the physics of complexity.

\subsection{One can not understand the quantum mechanics, but one can...}
This text is in about that one can understand the quantum mechanics. At least because that the basic principles of quantum mechanics, so paradoxical and counter-intuitive, and ``breaking down all classical physics''', reveals themselves already in the classical physics. And it happens already in relatively simple physical systems if one asks them right questions.
	
Quantum \textit{theory}	arose largely thanks to the questions that the nature asked for two previous decades in the language of classical physics \cite{EvoQuant}. E.g., E. Schr\"odinger suggested his model of atom (including the quantisation of angular momenta) on the basis on spherical waves \cite{Sch,Polak}, L.De-Broglie used in his work on ``matter waves'' variational principles of classical mechanics \cite{DeB,Polak}, and W.Heisenberg formulated his matrix mechanics in the grounds of Kramers--Kronig--Brilluen model of light propagation in the optically dense media \cite{Hei,HeiUFN}. 

In a sense, the ``quantum world'' still obeys the lows of the ``classical word'' -- but not the simplest ones. But this natural and essential point is for some reason almost absent in quantum mechanics courses (e.g., \cite{LL3,FLF8,LvQM}). The goal of out mini-course is to partially fix this omission.

\

The mini-course is based on the three popular stories.	

\
	
In sec.\ref{sec:RBLight}, we consider an extremely simplified model of (incoherent) transition of light through a medium. We use this basic example to follow the logic of Heisenberg and ``discover'' the matrix (operator) mechanic. Then we introduce a linear space of quantum states and ``calculate'' the evolution operator as a path sum. We also discuss the relation of our model to its physical prototype, and to a model of atom.
	
\
	
In sec.\ref{sec:mulSch}, we explore asymptotics (dependence on direction to the system and on distance from it when the distance is much more than size of the system) for the potential of system of electric charges fixed in a bounded space area. We present the solution in a nice form of (electrostatic) \textit{multipole expansion} and see a straightforward analogy between a ``classical atom'' (a ``cake'' of electric charges) and a ``quantum atom'' in Schr\"odinger model. Discussing conceptual grounds for this analogy, we come to the quantisation formalism for angular momenta and to the connection of system's symmetries with conservation laws in classical and quantum mechanics. Finally, we formulate fundamental differences of the ``classical'' and ``quantum'' angular momenta and discuss a little the problem of simultaneous (im)measurability of  quantum observables. 
                            
\
	
We devote sec.\ref{sec:hidsymm} to the main potential of our Universe, the $1/r$ potential, which enters Coulomb and Newton laws. We will comprehend this unique hidden symmetry of the potential to which we owe closed orbits of planets, degenerate spectrum of hydrogen, and celebrated experiments testing General relativity and Quantum electrodynamics. Here we need a bit more of standard formalism of classical and quantum mechanics. We do not mean to present this formalism sequentially here. Instead that, we give its main ideas and necessary references. Apart from that, a straightforward check of some key facts requires for rather cumbersome computations, which we added to the Appendix.

\subsection{Principles of quantum mechanics\label{sec:qprinz}}
Now we describe, in the most general terms, main ideas that underlie the physical approach to complicated, (partially) indeterminate systems. Essentially these are the standard postulates of quantum mechanics \cite{LL3,LvQM}, presented in spirit of \cite{FLF8,FeyInt}. We will see how these ideas work for particular physical systems in the consequent sections (which form the main content of the mini-course).

\paragraph{\S1. Indeterminism as microdeterminism.} 
For indeterminate system, a set of all physical quantities that \textit{principally} can be measured in a given system in a given moment of time does \textit{not} determine the set of values of the same quantities in a subsequent moment of time. I.e., a physical state of the system does not \textit{uniquely determine} the evolution of the state in time. Instead that, various \textit{evolution paths} are possible, with the following properties being supposed.  
\begin{enumerate}
\item{\textbf{Completeness.} The set of all evolution paths for a given system from a given state is a property of the system\footnote{This is the case in ``regular'' quantum mechanics, i.e., in a (quasi)stationary theory of a given numbers of particles. As soon as one must consider something principally non-stationary, e.g., a radiation that \textit{emerges} in the system, or particles birth, annihilation and transformation one to another, this Completeness property can not be preserved in the same form. As one tries to adopt it to these new cases, completely different science emerges, which is a quantum field theory \cite{FeyQED}.}.}
\item{\textbf{Composition.} Each evolution path from moment of time 1 to moment of time 2 can be followed by any evolution path from moment 2 to moment 3.}
\item{\textbf{Superposition.} If one substitutes in the above composite evolution path from time moment 1 to time moment 3 the evolution path from 1 to 2 OR from 2 to 3 -- one has another path from 1 to 3.}
\item{\textbf{Uncertainty.} As system evolves from the initial to the final moment of time, one can \textit{in principle} measure only the quantities that are the same \textit{for all} evolution paths\footnote{In formal quantum mechanics, the corresponding quantities are called invariants on the space of paths \cite{FeyInt}}.}
\end{enumerate}
An example of a measurable quantity in item 4 is the total number of paths, which is trivially independent of choice of a path.
Postulates 1)--3) can be embedded in the one
\begin{center}
\textbf{Superposition principle:}\hspace{2cm}
\{All evolution ways from time 1 to time 3\}= \\ 
$\bigoplus$
\{All evolution ways from time 1 to time 2\}$\times$
\{All evolution ways from time 2 to time 3\}.
\end{center}
Note that the above formula contains \textit{direct} product and \textit{direct} sum. Now we talk just about plain enumeration of all paths. The formula gets a physical sense only after defining physical quantities from item 4, i.e. of the quantities that can be \textit{in principle} measured according to the formula\footnote{In formal quantum theory, this is called a measure on the space of evolution paths\cite{FeyInt}.}.

\paragraph{\S2. Measuring a quantity as modifying if the system.}
For systems of our interest, a measurement of a physical quantity is a process of \textit{irremovable} modification of state of the system. I.e., a measurement is a particular case of evolution of the system. Hence, a measurement possesses all properties from \S1.
\begin{enumerate}
\item{A measurement is \textit{irremovable} impact on the system that causes evolution of the system determined by the measured quantity.}
\item{In a process of measurement, one can \textit{in principle} determine only the \textit{observables}, i.e. the quantities that are the same for all paths of the evolution associated with the measurement.}
\item{The observable takes a definite value or \textit{is conserved} if the measurements in different evolution moments \textit{always} give the same value.}
\item{The two observables can take definite values simultaneously or are \textit{simultaneously measurable}, if measurements of one quantity in various times of measuring of the other quantity \textit{always} give the same value.}
\end{enumerate}

\

Given preliminary formulations of the quantum principles, let's look how they work for particular physical systems.

\section{From light scattering to matrix mechanics\label{sec:RBLight}}	
\subsection{Physical phenomenon}
At the end of February 1928 Raman and Krishnan in Kalkutta University, and Landsberg and Mandelstam at Moscow State University, observed curious optical phenomenon. Namely, the spectrum of light scattered on quartz or Iceland spar crystal contained, together with each line of the incident light, several additional lines. The additional lines always came in pares, ``red'' on the left and (much more pail) ``blue'' on the right of the principle line. Moreover, the ``red''--principle and principle--``blue'' frequency differences coincided with each other and depended only on the media, but not on the frequency of the principle line. The phenomenon was similar to previously known Mandelstam--Brilluen effect, which is Doppler shift of the incident light wave length in a crystal. However, in a new phenomenon the split of the ``red-blue'' doublet was much more than Doppler split. This forced them to search for a principally new phenomenon of the new split, which was called a \textit{combination scattering} \cite{Siv4,CombScatt}.

Combination scattering was predicted already in 1910-s by nearly all founders of quantum mechanics. In classical theory of light transition through a media, the incident light as an electro-magnetic wave induces the oscillations of valent electrons of the media. They radiate then secondary electro-magnetic waves, which are observed as a light passed through the media (or scattered by the media). Moreover, the secondary waves are coherent to the primary one, i.e., they have the same phase and frequency. Changing of the frequency (and of the phase) is then possible as modulation when the secondary wave is added to the thermal oscillations of the molecules or crystal bounds of the media. The classical theory of the combination scattering predicts the very presence of the ``red--blue'' doublet, but is majorally mistaken in details (e.g., it does not predict that the ``blue'' line is weaker than the ``red'' one). Principally different possibilities are opened by quantum theory, with its universal view on various phenomena associated with a light transition through a media.

We consider a toy model of light scattering through the media, which is similar to the combination scattering. The model is oversimplified compared to realistic experiment, but is extremely clear in demonstration of principles behind constructing a quantum model. 

\subsection{Minimum quantum model\label{sec:combscatt}}
\begin{figure}[h!]
    \centering

     \includegraphics[width=1\textwidth,height=4cm,keepaspectratio]{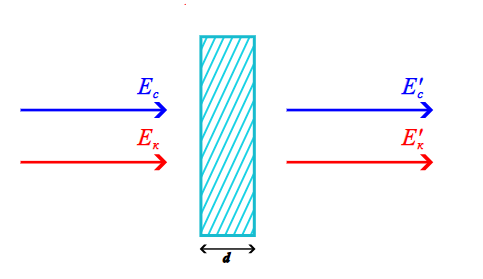}
    \caption{\footnotesize A toy model of light scattering.}
    \label{fig:RBscatt}

\end{figure}

Now consider transition of a beam of ``red'' and ``blue'' rays (below we omit the quotes implying the same) through a media. Split the media into several consequent layers. Suppose that on the boundary of each layer each ray can either change from red to blue or vice versa, or it may stay the same. The colours do not mix inside a layer (see Fig.\ref{fig:RBscatt}). 

A quantum model of the phenomenon is as follows.
\begin{enumerate}
    \item{The incident beam is composed of red and blue spectral components of intensities $I_r$ and $I_b$. Hence, the energies of the red and blue components that are transferred through the cross section with the area $S$ for the time of light transition through the $d$ thick layer are equal $E_1^r=I_rSd/c$, $E_1^r=I_rSd/c$, where $c$ is the light velocity.}
    \item{In quantum theory the spectrum components contain $N^r_1=E_1^r/(\hbar\omega_r)$ red and $N^b_1=E_1^b/\hbar\omega_b$ blue quanta.}
    \item{Each layer contains $N_r=n_rSd$ molecules in the ground state and  $N_b=n_bSd$ molecules in the excited state, whose energies differ for $\Delta_{br}$, where $n_r$ and $n_b$ are the corresponding volume densities, which are the properties of the medium.}
    \item{Each blue quantum excites each molecule in the ground state with probability $p_{rb}$, loosing the energy $\Delta_{br}$ and decreasing the frequency by $\Delta_{br}/\hbar$. Similarly, each red quantum relaxes each exited molecule with the probability $p_{br}$, acquiring the energy  $\Delta_{br}$ and increasing the frequency by $\Delta_{br}/\hbar$.}
    \item{If the spectral frequencies of the incident beam are equal $(\omega_b-\omega_r)\hbar=\Delta$, then
    an interaction with media molecules leads only to redistribution of the light quanta between the two initial spectral components, and no new spectral components appear.}
    \item{As a result of transition the of light through the layer (12), $N^r_1$ red quanta relax $N_bN^r_1p_{rb}$ molecules and become blue, and $N^b_1$ blue quanta excites $N_rN^b_1p_{br}$ molecules and become red. Energy of the red spectral component that changes by 
    $E^r_2 - E^r_1 = -N_bp_{rb}N^r_1\hbar\omega_r+N_rp_{br}N^b_1\hbar\omega_b=-N_bp_{rb}E^r_1+N_rp_{br}E^b_1$,
    and similarly for the blue component.}
    \item{Dividing the both parts of the equalities by $Sd/c$, write down the expressions for the new intensities:}
\end{enumerate}

\begin{equation}
    \begin{array}{rcrcr}
     I^r_2&=&I^r_1+dS\left(-n_bp_{rb}I^r_1+n_rp_{br}I^b_1\right)&=&W^{rr}_{21}I^r_1+W^{rb}_{21}I^b_1\\\\
     I^b_2&=&I^b_1+dS\left(-n_rp_{br}I^b_1+n_bp_{rb}I^r_1\right)&=&W^{br}_{21}I^r_1+W^{bb}_{21}I^b_1
    \end{array}.\label{transeq}
\end{equation}

\subsection{Matrix mechanics}
If one applies the described model to the sequent transition of the beam through the two layers, one gets (i.e., for the red component):
\begin{equation}
\begin{array}{r}
     I^r_3=W^{rr}_{32}I^r_2+W^{rb}_{32}I^b_2=
     W^{rr}_{32}\left(W^{rr}_{21}I^r_1+W^{rb}_{21}I^b_1\right)+W^{rb}_{32}\left(W^{br}_{21}I^r_1+W^{bb}_{21}I^b_1\right)=\\\\=
     \left(W^{rr}_{32}W^{rr}_{21}+W^{rb}_{32}W^{br}_{21}\right)I^r_1+\left(W^{rr}_{32}W^{rb}_{21}+W^{rb}_{32}W^{bb}_{21}\right)I^b_1\equiv 
     W^{rr}_{31}I^r_3+W^{rb}_{31}I^b_3\end{array}.\label{transcomp}    
\end{equation}
This is the same law as for each layer separately, but with new coefficients $W$. The latter identity in (\ref{transcomp}) imply that
the coloured components of $W_{21}$, $W_{32}$, and $W_{31}$ satisfy the relations in those one can find out the matrix multiplication. I.e., one can rewrite (\ref{transcomp}) as
\begin{equation}
  \left(\begin{array}{cccc}
I_3^r\\
I_3^b
\end{array}\right) =
\left(\begin{array}{cccc}
W_{32}^{rr} & W_{32}^{rb}\\
W_{32}^{br} & W_{32}^{bb}
\end{array}\right)
\left(\begin{array}{cccc}
W_{21}^{rr} & W_{21}^{rb}\\
W_{21}^{br} & W_{21}^{bb}
\end{array}\right)
\left(\begin{array}{cccc}
I_1^r\\
I_1^b
\end{array}\right)=
\left(\begin{array}{cccc}
W_{31}^{rr} & W_{31}^{rb}\\
W_{31}^{br} & W_{31}^{bb}
\end{array}\right)
\left(\begin{array}{cccc}
I_1^r\\
I_1^b
\end{array}\right).
\label{transmat}
\end{equation}
In other words, a state that is a vector of the spectral components $\mathbf{I}$ is subjected to the linear operator $\hat W$,
\begin{equation}
\mathbf{I}_3=\hat W_{31}\mathbf{I}_1=\hat W_{32}\hat W_{21}\mathbf{I}_1,\label{transop}
\end{equation}
that defines the evolution of the state as the light beam passes through sequent layers of the media. Each of relation (\ref{transcomp}), (\ref{transmat}), (\ref{transop}) implies that the operator $\hat W$ by construction  obeys the \textit{superposition principle}: 
\begin{equation}\hat W_{31}=\hat W_{32}\hat W_{21}.\label{supop}\end{equation}

\subsection{A sum over paths}
The expanded form of (\ref{transcomp}) related superposition principle (\ref{supop}) with the \textbf{Superposition principle} from sec.\ref{sec:qprinz}:
\begin{equation}
    \begin{array}{cccc}
   &\multicolumn{3}{c}{\text{Any paths for} a,b,c\in \{r,b\}:}\\\\
         \sum\limits_b&W^{ab}_{32}&*W^{bc}_{21}&=W^{ac}\\
         \text{Sum over all}&\text{from state $b$ in section 2}&\text{from state $a$ in section 1}&
    =\text{from state $a$ at point 1}\\
         \text{states in section 2}&\text{to state $c$ in section 3}&\text{to state $b$ in section 2}&\text{to state $c$ in section 3}
    \label{pathcomp}\end{array}.
\end{equation}
In notation (\ref{pathcomp}), we mean by a path a transition between two spectral components of the state $\mathbf{I}$ as the light beam passes a layer of the medium (i.e., transition between $r$  component  in section 1 and $r$ component in section 2). A measure of this path is the corresponding matrix element of the evolution operator $\hat W$ ($W^{rr}_{21}$ in the above example), and a physical sense of this measure is revealed by (\ref{transeq}). 

One can take a partition of a given layer into any number of intermediate layers. Applying then (\ref{pathcomp}) sequently time after time, one can write down the same transition as a sum over any number of intermediate states,
\begin{equation}
W^{ca}_{m,1}=\sum_{b_1,\ldots,b_k}W_{m,k+1}^{cb_k}\ldots W_{32}^{b_2b_1}W_{21}^{b_1a}.\label{pathint}
\end{equation}
Here $b_1,\ldots,b_k$ are the colour indices that take values $r$ or $b$, while $1$ and $m$ are the fixed initial and final cross sections, and positions of the intermediate cross sections $1,\ldots,k$ depend on the number of iterations $k$ and on the partition, as well as the corresponding $\hat W_{b_{i+1}b_i}$. The sense of the formula is that one can define the same $\hat W$ as a function of two sections (initial and final one), so that for any partition (\ref{pathint}) is valid with the proper $\hat W_{b_{i+1}b_i}$. This is a curious and highly non-trivial property of the evolution operator. As we saw above, this property formally follows just from linearity of our model. Yet already several quantum postulates are essentially  built in this property. In turn, each postulate has behind it a certain (and not always valid) physical assumption.

\begin{figure}[h!]
    \centering

     \includegraphics[width=1\textwidth,height=4cm,keepaspectratio]{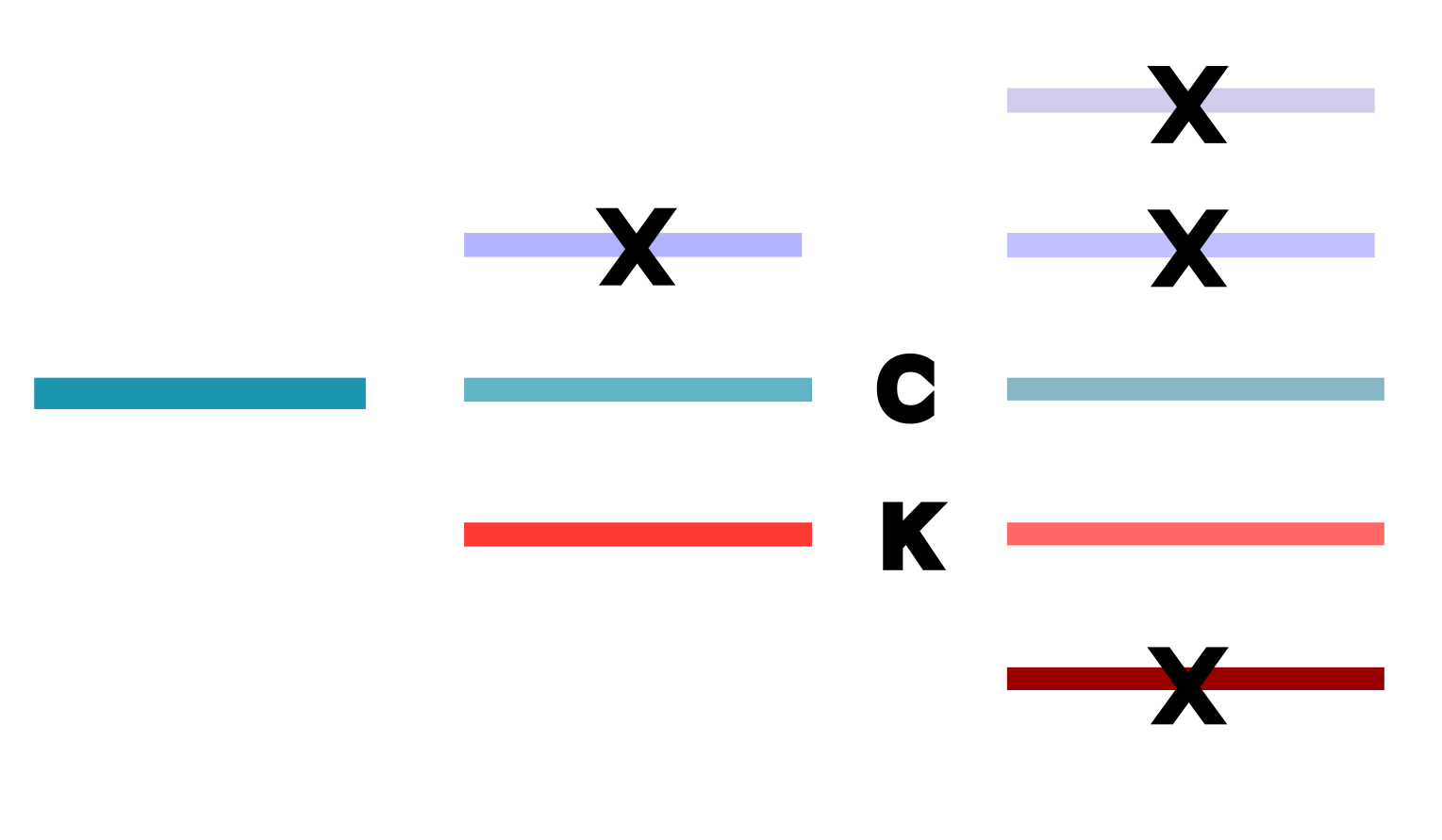}
    \caption{\footnotesize Assumptions of the two-colour model.}
    \label{fig:RBLight}

\end{figure}

\subsection{Physics of quantum postulates}
Particular computable form (\ref{pathcomp}) of formal \textbf{Superposition principle} contains postulates from sec.\ref{sec:qprinz} in the following form.

\begin{enumerate} 
\item{\textbf{Completeness.} Spectral indices $a,b,c$ always run values $r$ and $b$ (and the sum is taken in the same limits). I.e., no new colours arise as the beam passes through a layer. In other words, no new quantum states arise during the evolution.

Strictly speakings, this postulate is wrong in the model of combination scattering from sec.\ref{sec:combscatt}. Indeed, each new layer may give rise to all new frequencies differing from the initial ones by $n\Delta/\hbar$, where $n$ is generally any integer. However, we can consider a model, which is rather fantastic from the standpoint of physical media, but very instructive for the analysis.

As we said, in physical media ratio of intensities of the ``more blue'' and ``more red'' components is much less than one. Imagine a medium where the same is true for the ratios of the ``more blue'' and initial components and of the initial and ``more red'' components, and the both ratios are of the same order $\delta$. Moreover, let the ratio of the initial and final red and blue components be of the same order. Then the relations of our two-colour model are can be presented as a part of real, many-colour model, namely as relation of the first order in $\delta$ (see fig.\ref{fig:RBLight}).}

\item{\textbf{Composition.} Intensities of the spectral components  $\mathbf{I}_k$ outgoing a sequent layer are considered as the incoming ones, and they obey the same evolution relations (although the matrix elements $\hat W_{k,k+1}$ may be new for each layer). Hence the evolution of the beam is \textit{local}.

This postulate follows just from the physical model from sec.\ref{sec:combscatt}. Namely, outgoing beam is composed of the same photons, as the incident one. Moreover, they interact with medium in the same way, and interaction of each photon with the molecules does not depend and doest not impact on properties of other photons, molecules, and media in common. This postulate would be violated in any slightly more complicated model, e.g., if other particles were radiated apart from photons, or if photons behaved as one whole (as in laser radiation), or a if a global process took media in the material (such as ferromagnetism, M\"ossbauer effect, superconductivity), ect.}

\item{\textbf{Superposition.} Contribution of intensities of each spectral components from all evolution paths are summed up. Moreover, change of any transition (red-to-red for red-to-blue or wise verse; blue-to-red for blue-to-blue or vice verse) always gives a new evolution way, i.e., red and blue components are always \textit{distinct}.

This postulate has a \textit{incoherency} of the combination scattering behind it: the secondary waves do not interfere with each other. Hence, their intensities, the waves themselves are added. It is different in the case of ordinary transfer of the light in the vacuum or in a medium, when Huygens--Fresnel principle is satisfied. In the latter case the secondary waves are coherent, and the passed beam is obtained as a result of interference of the secondary waves.}

\item{\textbf{Uncertainty.} The formula contains the \textit{sum over all intermediate states} ($r$ and $b$). I.e., it is possible to predict evolution of the red or blue component of the beam  is \textit{in principle} possible only by considering the \textit{coevolution} of the both components. In other words, it is \textit{principally} impossible to predict the evolution of the red (blue) component by considering only the red (blue) component.

This is the postulate that makes out model ``quantum'' -- and it needs a separate comment.}
\end{enumerate}

\subsection{Observables and unobservables}
The above \textbf{uncertainty} reflects the physical principle: one can measure only the total numbers of transitions of each kind on a layer boundary -- but this measurement \textit{in principle} does not allow one to follow the ways of these transitions inside the layer. Only the total numbers of the transitions are measured via intensities of spectral components of the outgoing beam (compared to those of the incident one).

\

Up to now the followed the statistical approach. Namely, we considered the matrix elements of $\hat W$ as \textit{numbers} of various spectral transitions for a big number of incident light quants. The same quantities can be considered as \textit{probabilities} of different spectral transitions meant as random events. From this standpoint, combinations of spectral indices enumerate all possible results. The composition property (the multiplication in the \textbf{Superposition principle}) then postulates the independence of spectral transitions in the consequent layers (like landing heads or tails in successive tosses). In turn, the superposition property (the addition in the \textbf{Superposition principle}) postulates the incompatibility of different transitions in one layer (like landing heads or tails in the same toss).

\subsection{Continuous evolution \label{sec:evoexp}}
Now we return to the \textbf{Superposition principle} in form (\ref{pathint}) with elements of $\hat W$ from (\ref{transeq}), where we set that $d$ tends to zero, and the number of layers $N$ tends to the infinity, so that the total thickness $dN=D$ is constant. Let for simplicity all layers have the same density. Then one can write (\ref{transcomp}) in the form
\begin{equation}
\hat W_{i+1,i}\equiv \hat W_{0;\frac{D}{N}}=\left(\begin{array}{cc}1&0\\0&1\end{array}\right)+
\textstyle{\frac{DS}{N}}\left(\begin{array}{cc}-n_bp_{rb}&n_rp_{br}\\
n_bp_{rb}&-n_rp_{br}\end{array}\right)\equiv \mathrm{Id}+\textstyle{\frac{DS}{N}}\hat w,
\label{infevo}\end{equation}
and (\ref{pathint}) is rewritten as
\begin{equation}
\hat W_{N,1}\equiv \hat W_{D;0}=\lim_{N\to\infty}\hat W^N_{0;\frac{D}{N}}=
\lim_{N\to\infty}\left(\mathrm{Id}+\textstyle{\frac{DS}{N}}\hat w\right)^N=\exp\left(DS\hat w\right)=
\sum_{k=0}^\infty \frac{1}{k!}\left(DS\hat w\right)^k.\label{evoexp}
\end{equation}
In the end of expression (\ref{evoexp}) we used the definition and a property of the matrix exponential \cite{MatExp}.
The operator $S\hat w$ is said to generate an \textit{infinitesimal evolution} in the evolution parameter $D$. Hence,
\begin{center}
\begin{tabular}{cp{15cm}}
$\bullet$&Writing down the evolution operator as a path sum (\ref{pathint}) allows one to rewrite it as the exponential of the operator that generates infinitesimal evolution multiplied by the evolution parameter -- (\ref{evoexp}). \\
$\bullet$&The both representations of the evolution operator are consequences of the \textbf{Superposition principle} (\ref{pathcomp}).
\end{tabular}
\end{center}

\subsection{Measurement and translation symmetry\label{sec:trans}}
According to sec.\ref{sec:qprinz}, a quantum measurement -- is a process of evolution of the system, and hence one can measure only a quantity that is the same for all evolution paths. Let's see, which quantities can be measured in the above evolution of a light beam. 
First, 
\begin{center}
\begin{tabular}{cp{15cm}}
$\bullet$&The total number of evolution paths at each step $N$ is the same for all paths.
\end{tabular}
\end{center}
In our problem generally $N=2$.
Second,
\begin{center}
\begin{tabular}{cp{15cm}}
$\bullet$&If there is just one evolution path for an initial state, then any function of this path is an observable for this state.\end{tabular}
\end{center}
I our problem, this is the case if the vector of initial intensities is an eigenvector of the infinitesimal evolution operator \cite{Gel}, i.e., (\ref{infevo}) becomes 
\begin{equation}
\frac{N}{DS}\left(\begin{array}{c}I_2^r-I_1^r\\I_2^r-I_1^b\end{array}\right)=\hat w
\left(\begin{array}{c}I_1^r\\I_1^b\end{array}\right)=
\lambda \left(\begin{array}{c}I_1^r\\I_1^b\end{array}\right)
\label{eigenw}\end{equation}
Then the observable for this state is the eigenvalue $\lambda$. 
In case of homogeneous media, i.e., in case of \textit{translation symmetry} along the beam propagation, one can also write relation (\ref{evoexp}), which takes the form
\begin{equation}
\left(\begin{array}{c}I_N^r\\I_N^b\end{array}\right)=
\hat W_{N,1}\left(\begin{array}{c}I_1^r\\I_1^b\end{array}\right)=
\exp(DS\lambda)\left(\begin{array}{c}I_1^r\\I_1^b\end{array}\right).\label{eigenW}    
\end{equation}
Moreover,
\begin{equation}
\frac{1}{S}\frac{\partial}{\partial D}\left(\begin{array}{c}I_N^r\\I_N^b\end{array}\right)=
\lambda\left(\begin{array}{c}I_N^r\\I_N^b\end{array}\right),\label{ddDlambda}    
\end{equation}
while for general state one can write
\begin{equation}
\frac{1}{S}\frac{\partial}{\partial D}\left(\begin{array}{c}I_N^r\\I_N^b\end{array}\right)=
\hat w\left(\begin{array}{c}I_N^r\\I_N^b\end{array}\right).\label{ddDw}    
\end{equation}
In this sence one can talk about the correspondence of the operators $\hat w$ and $\frac{\partial }{\partial D}$, i.e.,
\begin{center}
\begin{tabular}{cp{15cm}}
$\bullet$&In the framework of the considered model of a light beam propagation, one can represent the infinitesimal evolution operator as the derivative operator along the beam propagation,
in the sence of relations (\ref{ddDw}).
\end{tabular}
\end{center}
As soon as the derivative operator by definition can be considered as the operator of the infinitesimal shift, one can say that the eigenvalue $\lambda$ for a state eigenvector, as well as the dimension of the vector (which is the number of paths for a general state at each evolution step), are \textit{invariants} of the translation symmetry.

We note also that if the beam propagates with the velocity $c$, than we can with the same success talk about the state evolution in time $t$ and about the state evolution  in the media thickness $ct$.

\subsection{Light scattering and the atom model}
Considering the light scattering in a media in the framework of a model similar to the above simple model, Kramers, Kronog and Heizenberg \cite{Hei, HeiUFN} came to the notion of the vector of atom states and to the matrix of transition between the states. The atom states are unobservable, but some matrix elements are observable via the thicknesses of spectral lines. Moreover, the transition matrix satisfies the \textbf{Superposition principle}, which Heizenberg used to derive his quantisation rules for various function of classical observables. This is how the foundations of the matrix quantum mechanics were laid.

\section{From electrostatics to Schr\"odinger equation\label{sec:mulSch}}	

\subsection{An atom with ``raisins'' and Bohr atom}
Naive classical model considers an atom as a ``bun with raisins'', where historically the ``bun'' was the positively charged nucleus, and the ``raisins'' were negatively charges electrons. 
Yet in Reserford experiments, atom behaved itself rather as a profiterole, with volume and full of ``holes'' electronic ``dough'' and with very compact nucleus in the centre as a ``filling''. Hence electrostatic model of an atom appeared, where an electron moves in an orbit in the electric field of the nucleus.
Unfortunately, such an atom can not be stable. But there are other problems as well.

As it became clear in the beginning of the XX century, it is \textit{principally} impossible to observe a position of an electron in the atom, as well as its movement in the orbit.
All we can observe \textit{in principle} describes describes state of the electron in another way, by \textit{quantum numbers}. The first example is a Mendeleev periodic table of elements. As we now know, the table reflects patterns of filling electron shells. Namely, the group number is related to the number of electrons on the valent shell, period number equals to the principle quantum number of the valent shell, and the element number equals to the total number of electrons. The observed spectral lines of atoms are also described by the principle quantum number (Lyman, Balmer, Paschen, ect. series) and orbital quantum number (sharp series ($s \rightarrow p$), principle series ($p \rightarrow s, p\rightarrow d$), diffuse series ($d \rightarrow s$)).

As an answer to these problems and observations, Niels Bohr suggests a new model of an atom, where both the atom stability and the description by quantum numbers (as \textit{the only possible} one) were \textit{postulated}. It seems that the Bohr model totally contradicts to the electrostatic one. But this is wrong, as Erwin Schr\'odinger demonstrated soon. Below we try to reproduce the logic of this story, as we understand it. We start from the classical model, like a hybrid of atom with ``raisins'' and atom with ``filling''. Namely, we put arbitrary charges of any sighs in the ``centre'' and one more test charge far from the ``centre''. In the following, we will see how this model is related to the real atom. But first we will see the formal analogy of our electrostatic model and Bohr atom.

\subsection{States of far system of charges\label{sec:mulstate}}
\begin{figure}[h!]
    \centering

     \includegraphics[width=1\textwidth,height=6cm,keepaspectratio]{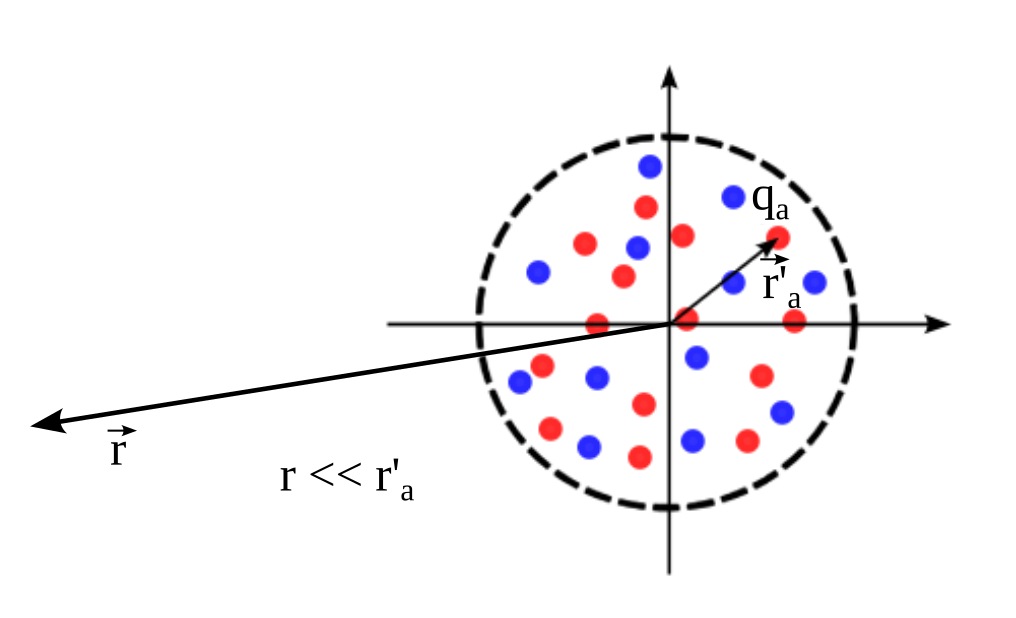}
    \caption{\footnotesize A system of charges, far from an observer.}
    \label{fig:multipole}

\end{figure}

\paragraph{Multipole expansion of the potential.}
Consider a system of charges (generally, or arbitrary magnitudes and signs)  (arbitrary) placed in a bounded area of space, which is far from the observer (i.e., the distances between charges are much less than the distance between a charge and the observer), see fig.\ref{fig:multipole}. Generally, electrostatic potential of the system of charges equals \cite{Jack}
\begin{equation}
\varphi(\vec{r})=\int\frac{\rho(\vec r\,')dV'}{\mid\vec r-\vec r\,'\mid},\label{LapGreen}
\end{equation}
where $\vec r$ -- is the radius vector of the observer w.r.t. an arbitrary fixed point inside the area of charges, and $\vec r\,'$ is the radius vector from the fixed point that runs over all volume $V'$ of the charged area. 
In our case $\mid\vec r\,'\mid\equiv r' <\!\!<\ \mid\vec r \mid \equiv r$ one can expand the integrand in (\ref{LapGreen}) into the Taylor series, denoting by $x_{i}$, $x'_i$ the components of $\vec r$,
$\vec r\,'$ and implying (everywhere below) \textbf{that the sum is taken over the repeated indices} from  1 to 3: 
\begin{equation} \label{GreenTay}
\frac{1}{\mid\vec r-\vec r\,'\mid}=\frac{1}{r}+\frac{1}{r^{3}}x_{i}x'_{i}+
\Big\{\frac{3}{2r^{5}}x_{i}x_{j}x'_{i}x'_{j}-\frac{1}{2r^{3}}x'_{i}x'_{i}\Big\}+
\Big\{\frac{3}{2r^5}x_jx'_jx'_ix'_i-\frac{5}{r^7}x_ix_jx_kx'_ix'_jx'_k\Big\}+
\frac{1}{r}\mathcal{O}\textstyle{\Big(\left(\frac{r'}{r}\right)^4\Big)},
\end{equation}
wherefrom
\begin{eqnarray}\label{IntTay}
\varphi(\vec r)=\frac{1}{r}\int\rho(\vec r\,')dV'+\frac{1}{r^2}\,n_{i}\int x'_{i}\rho(\vec r\,')dV'
+\frac{1}{2r^{3}}\left\{3n_{i}n_{j}-\delta_{ij}\right\}\int x'_{i}x'_{j}\rho(\vec r\,')dV'
+\nonumber\\+
\frac{1}{2r^{4}}\left\{5n_{i}n_{j}n_k-\delta_{ij}n_k-\delta_{ik}n_j-\delta_{jk}n_i\right\}\int x'_{i}x'_{j}x'_k\rho(\vec r\,')dV'
+\cdots,
\end{eqnarray}
where $\vec n$ is the unit vector, directed from a given point inside the system of charges towards the observer; $\vec n=(\sin\theta\cos\varphi, \sin\theta\sin\varphi,\cos\theta)$ in the spherical coordinates.
Hence, one can write down expansion (\ref{IntTay}), separating in each term by separating the dependences on the distance from (a fixed point inside) the system to the observer $r$, on the unit vector $\vec n$ from the observer towards the  system (or on the spherical angles $(\theta,\varphi)$), and on the charge density $\rho(\vec r\,')$,
\begin{eqnarray}\label{MultTay}
\varphi(\vec r)=\frac{Q}{r}+\frac{n_iD_i[\rho( \vec r\,')]}{r^2}+\frac{\Omega_{ij}(\vec n)K_{ij}[\rho (\vec r\,')]}{r^3}+\frac{\Omega_{ijk}(\vec n)O_{ijk}[\rho( \vec r\,')]}{r^4}
+\ldots\end{eqnarray}
Generally, the expansion of  (\ref{IntTay}) has a similar form, 
\begin{eqnarray}\label{MulPot}
\varphi(\vec r)=\sum_{l=0}^{\infty}\frac{(-1)^{l}}{l!}\int\limits\rho(\vec r\,')x'_{i_1}\ldots x'_{i_1}
\frac{\partial}{\partial x'_{i_l}}\ldots\frac{\partial}{\partial x'_{i_l}}\frac{1}{\mid \vec r-\vec r\,'\mid}\Bigg|_{r'=0}\!\!\!\!dV'=\nonumber\\=
\sum_{l=0}^{\infty}\frac{(-1)^{l}}{l!\,r^{l+1}}\Big[r^{l+1}\frac{\partial}{\partial x_{i_l}}\ldots\frac{\partial}{\partial x_{i_l}}
\frac{1}{ r}\Big]\int\limits\rho(\vec r\,')x'_{i_1}\ldots x'_{i_l}dV'=
\sum_{l=0}^{\infty}\frac{(-1)^{l}}{l!}\frac{\Omega_{i_1\ldots i_l}(\theta,\varphi)}{r^{l+1}}M_{i_1\ldots i_l},
\end{eqnarray} 
where $M_{i_1\ldots i_n}$ depends only on the properties of the system of charges, and the multiplier of it depends only on the observables coordinates. Moreover, $\Omega_{i_1\ldots i_n}$ depends only on the unit vector $\vec n$ and thus only on the angular coordinates $\theta,\varphi$, as an invariant of the change $r\to \kappa r$.

Hence, general potential (\ref{LapGreen}) is expanded far from the system of charges into the sum of potentials with much simpler dependence both on the position of the system of charges w.r.t. the observer and on the charge density function inside the system (because each term contains only the integral of the product of the charge density and the coordinates, i.e., a \textit{moment}of given order of the density function, but not the set of the function values in all points).

Expansion terms (\ref{MultTay},\ref{MulPot}) can be seen as contributions into the total potential of potentials of some subsystems, into which we imaginary divide the main system of charges. Such subsystems are called \textit{multipoles}, and series (\ref{MultTay},\ref{MulPot}) is called electrostatic multipole expansion.

Hence we just applied one of the ground methods of theoretical physics. Namely, we expanded a complicated \textit{liner} system into a sum of much more simple subsystems. But it is much more important for us that we tuned in this way from the initial classical system to its space of states.

\paragraph{Multipole states.}
In ``quantum'' language, expressions (\ref{MultTay},\ref{MulPot}) mean: the system of charges with potential (\ref{LapGreen}) is for a far observable in a \textit{mixed state}. I.e., it is composed of subsystems in \textit{pure} states, which are \textit{multipoles}. More precisely, it is not reasonable to talk about dividing a system into subsystems in quantum mechanics, because our knowledge about construction of the system is \textit{principally} restricted. Instead that, one can interpret (\ref{MultTay},\ref{MulPot}) so that a mixed \textit{state} of the system is a \textit{sum of states}-multipoles.

Now consider each multipole state in more details. To do that, we write down the explicit expressions for quantities that describe the system of charges in (\ref{MultTay}) and are called \textit{multipole moments}, in case of a set of point-like charges, i.e., $\rho(\vec r\, ')=\sum_aq_q\delta(\vec r\, '-\vec r_a')$. One can read the simplest form of the multipole moments just from (\ref{IntTay}),
$Q=\sum_aq_a$,  $D_i=\sum_a q_ax'_{a,i}$, $K_{ij}=\frac{1}{2}\sum_a q_ax'_{a,i}x'_{a,j}$,  
$O_{ijk}=\frac{1}{2}\sum_a q_ax'_{a,i}x'_{a,j}x'_{a,k}$.
However, the multipole moments in (\ref{IntTay}) are defined not uniquely, but up to the change
\begin{equation}
    K_{ij}\to K_{ij}+G_{ij}:\ \ \Omega_{ij}G_{ij}=0\ \ \Rightarrow\ \ G_{ij}=\kappa\delta_{ij},\label{GaugeMult}
\end{equation}
and similarly for higher orders. Starting from $K$, the canonic expressions for the multipole moments differ from the naive one, so that the first multipole moments
($Q$ is the charge, $D$, $K$, $O$ are dipole, quadrupole, oktupole moments of the system) are
\begin{eqnarray}
    Q=\sum_aq_a,\  D_i=\sum_a q_ax'_{a,i},\  K_{ij}=\frac{1}{2}\sum_a q_a\big\{x'_{a,i}x'_{a,j}-\frac{1}{3}x'_{a,k}x'_{a,k}\delta_{ij}\big\},\nonumber\\
O_{ijk}=\frac{1}{2}\sum_a q_a\big\{x'_{a,i}x'_{a,j}x'_{a,k}-
\frac{1}{5}\delta_{ij}x'_{a,m}x'_{a,m}x'_{a,k}-
\frac{1}{5}\delta_{ik}x'_{a,m}x'_{a,m}x'_{a,j}-
\frac{1}{5}\delta_{jk}x'_{a,m}x'_{a,m}x'_{a,i}
\big\}.\label{MulTens}
\end{eqnarray}
Generally, the multipole tensors are constructed so that each tensor is fully symmetric under permutations of indices and the trace in any pair of indices is zero.
In particular, (\ref{MulTens}) implies that $K_{ij}=K_{ji}$ and $K_{ii}=0$, and $O_{ijk}=O_{jik}=O_{kij}=O_{ikj}=O_{kij}=O_{ikj}$ and $O_{iij}=0$.

In this definition of the multipole moments, the numerators in (\ref{IntTay}) are symmetric w.r.t. $\vec r$, $\vec r\, '$. Yet definition (\ref{MulTens}) has a much more profound sense, which we see below. 

\subsection{Multipole symmetries and quantum numbers\label{sec:symmult}}
\begin{figure}[h!]
    \centering

     \includegraphics[width=1\textwidth,height=4cm,keepaspectratio]{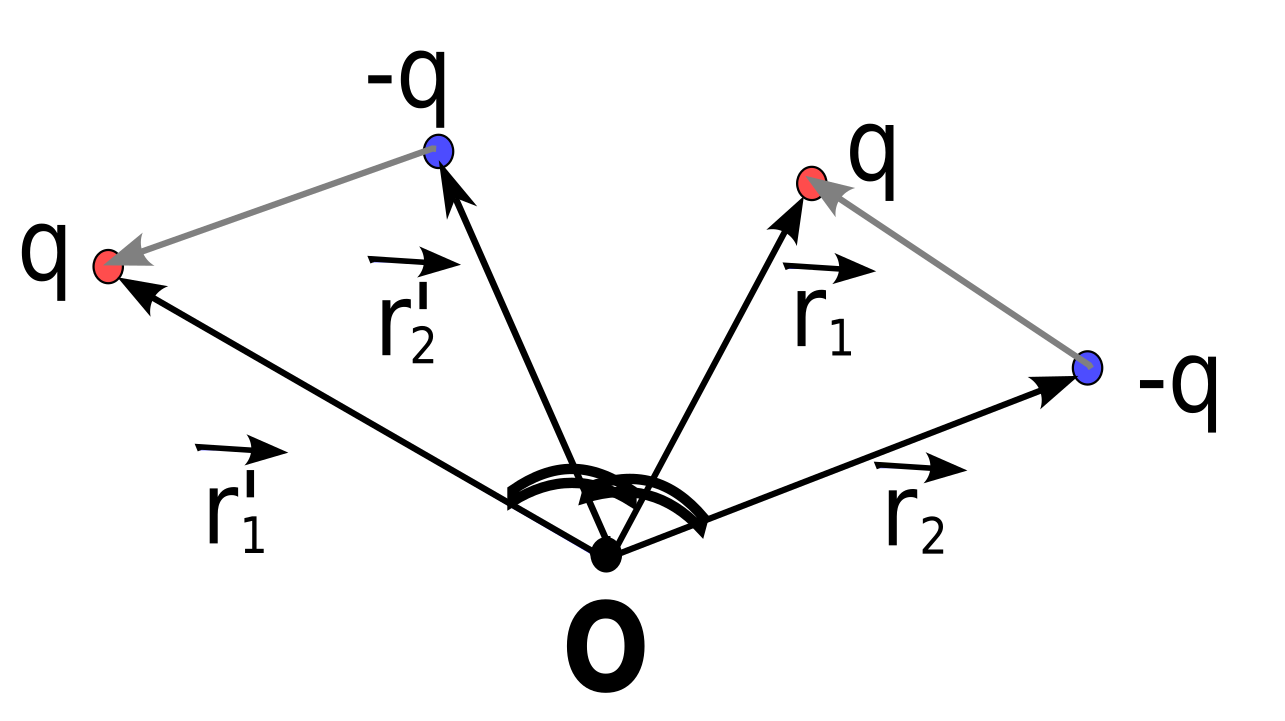}
    \caption{\footnotesize The simplest dipole and the result of its rotation as a whole.}
    \label{fig:dipole_rot}

\end{figure}

\paragraph{Rotation symmetry.}
Electrostatic potential is a scalar, i.e., a quantity that is independent of choice of coordinate system. Moreover, a contribution of each order in $\frac{r'}{r}$ in (\ref{IntTay},\ref{MulPot}) is also a scalar, because finite transformations can not mix these terms, and hence each term is invariant by itself. The total electric charge is also a scalar. The quantities $n_i$ and $D_i$ are already vector components. In particular, under rotation of the coordinate system\footnote{Instead of rotation of the coordinate system, one can talk about rotation of system of charges as a whole (see Fig.\ref{fig:dipole_rot})} by a matrix $U$ they become $U_{ij}n_j$ and $U_{ij}D_j$, and due to orthogonality of the rotation matrix ($U_{ij}U_{ik}=\delta_{jk}$) the scalar product $D_in_i=U_{ij}D_jU_{ik}n_k$ is preserved. The angular functions $\Omega$ in (\ref{IntTay},\ref{MultTay}), as well as the multipole moments 
$K$, $O$ in (\ref{IntTay}) and $M$ in (\ref{MultTay}) have similar properties. Namely, under rotation of the coordinate system by a matrix $U$ they transform as
\begin{equation}
    \Omega_{i_1\ldots i_n}\ \to\ \ U_{i_1j_1}\ldots U_{i_nj_n}\Omega_{j_1\ldots j_n},\ \ 
    M_{i_1\ldots i_n}\ \to\ \ U_{i_1j_1}\ldots U_{i_nj_n}M_{j_1\ldots j_n}.\label{MultTrans}
\end{equation}
In particular, $K_{i_1i_2}$ becomes $U_{i_1j_1}U_{i_2j_2}K_{j_1j_2}$ and $O_{i_1i_2i_3}$ becomes $U_{i_1j_1}U_{i_2j_2}U_{i_3j_3}O_{j_1j_2j_3}$.
I.e., by definition the quantities $\Omega$ and $M$ are by definition rank $n$ \textit{tensors}\footnote{Here we consider only orthogonal transformations of the Euclidean space (rotations), and hence we do not make difference between contra- and covariant tensors (super- and subscripts) \cite{Gel}.} \cite{Gel}. In particular, the charge $Q$ is a zero rank tensor or a scalar; dipole moment $D$ is a rank 1 tensor or vector; quadrupole moment $K$ is a rank 2 tensor, which can be presented as a $3\times 3$ matrix; oktupole moment $O$ is a rank 3 tensor, which can be presented as a $3\times 3\times 3$ ``cube''.

It follows from the above that the rotation group acts on a linear space of multiploles of the given order.
A ``group'' here means just that a composition of rotations is a rotation and that a rotation has the inverse one whose composition with the original one is the unity transformation. 
``Action'' of the rotation group element on a multipole means the rotation of the coordinate system with the corresponding transformation of the multipole tensor components. This operation gives components of another multipole tensor, because to rotation a coordinate system is equivalent to the inverse rotation of the multipole. In this sense the rotation group ``acts on the space of multipoles''. Finally, ``a linear space of multipoles'' means that a linear combination of multipole tensors is a multipole tensor, due to the electrostatic superposition principle. The corresponding system of charges is a composed of the two original systems with all charges rescaled by the coefficients of the linear combination.

In the above sense, one can say that
\begin{center}
\begin{tabular}{c|p{15cm}}
    $\bullet$& The linear space of multipoles of a given order is a representation space of the rotation group.
\end{tabular}
\end{center}
``Just'' this is enough to apply to description of the various multipoles, or of the state space of a system of charges in out interpretation, powerful tools of \textit{representation theory}.

\paragraph{Multipole space.}
Now it is time to discuss, why it is so important that the multipole tensors are symmetric and traceless (see sec.\ref{sec:mulstate}, at the end). If follows just from (\ref{MultTrans}) that a rotation of the coordinate system preserves the full symmetry under permutations of the indices. Moreover, the full symmetry of a tensor is preserved by a taking general linear combination of such tensors, hence they form a liner space. I.e., the space of fully symmetric tensors is an \textit{invariant space} of the rotation group. It follows also from (\ref{MultTrans}) that a rotation preserves the trace in any pair of indices, in particular, the zero trace,
\begin{equation}
    M_{i_1i_1i_3\ldots i_n}=0\ \Rightarrow\ \ 
    \underbrace{U_{i_1j_1}U_{i_2j_1}}_{\delta_{i_1i_2}}U_{i_3j_3}\ldots U_{i_nj_n}M_{j_1j_2j_3\ldots j_n}=
    M_{i_1i_1i_3\ldots i_n}=0,\label{MultTrless}
\end{equation}
The special about the zero trace is that its value is preserved by a general linear combination, unlike a non-zero trace. Hence, one can make a more precise statement, than the above one,
\begin{center}
\begin{tabular}{cp{15cm}}
      $\bullet$& The space of fully symmetric traceless tensors of a given rank is an invariant space of the rotation group.
\end{tabular}
\end{center}
I.e., this space, as well as the space of all rank $n$ tensors $n$, is a representation space of the rotation group. Moreover, one can demonstrate that any symmetric traceless tensor can be turned into any such tensor by a rotation, i.e. that the corresponding representation is \textit{irreducible} \cite{Vil}. One can apply similar arguments to the tensors $\Omega_{{i_{1}i_{2}}_{\cdots}i_{n}}$, which determine that angular dependence of the multipole potential. I.e., they also for a space of an irreducible representation of the rotation group for each $n$. Moreover,
\begin{center}
\begin{tabular}{cp{15cm}}
      $\bullet$&
The spaces of multipole tensors $M_{{i_{1}i_{2}}_{\cdots}i_{n}}$ and of angular dependence tensors $\Omega_{{i_{1}i_{2}}_{\cdots}i_{n}}$ can be equally considered as a space state of a system of charges for a far observable, \textit{instead of} the charge density function $\rho(x)$ at each point or the values and positions of each charge $\{q_a,x_a\}$.
\end{tabular}
\end{center}

Then by definition,
\begin{center}
\begin{tabular}{cp{15cm}}
     $\bullet$&\textbf{The state space of a system of charges} for a far observable, with the potential depending on the size of the system $r^{\prime}$ and on the distance to the observable $r$ by the law $\frac{r'^n}{r^{n+1}}$, for an integer $n$,\textbf{is a space of an irreducible representation of the rotation group}.
\end{tabular}
\end{center}
Representation theory allows one to determine the dimensions of all spaces of irreducible representations of the rotation group and to find in each space ``a good'' basis. 
However, before discussing a general approach to the problem, we solve it ``by hands'' and look at the answer in particular cases.

\paragraph{How many different multipoles there are?}
Recall the multipole tensors $M$ are by construction fully symmetric and traceless (generally $M_{i_1\ldots j\ldots j\ldots i_k}=0$, but due to the full symmetry it is enough to set $M_{jj i_3\ldots i_k}=0$), and that they form a linear space. Now we find its dimension. The dimension of space of the fully symmetric tensors in the 3d space equals to the number of various sets of $n$ numbers that take values 1, 2, and 3, which is $\frac{(n+1)(n+2)}{2}$.
The traceless condition means vanishing of a fully symmetric rank $n-2$ tensor that is the contraction or the original tensor, say, in the first two scripts. Hence, the dimension of space of the tensors that satisfy the both conditions equals
\begin{equation}\label{MultDim}
    \dim{P}=\frac{(n+1)(n+2)}{2}-\frac{(n-2+1)(n-2+2)}{2}=2n+1.
\end{equation}
This is the dimension of the space of the order $n$ multipole states.


Enumerate the $2n+1$ basis multipoles in the first orders explicitly.
\begin{equation}
    \begin{array}{p{1.5cm}cc}
             &\text{Form of the tensor}&\text{Basis tensors}   \\
            Charge&Q&1\\\\
            Dipole&D_i=\big(D_1,D_2,D_3\big)
            \begin{array}{c}
                  (1,0,0)\\
                  (0,1,0)\\
                  (0,0,1)
            \end{array}
               \\\\
            Quandrupole&K_{ij}=\left(\begin{array}{ccc}
                 K_{11}&K_{12}&K_{13}\\
                 K_{12}&K_{12}&K_{23}\\
                 K_{13}&K_{23}&-K_{11}-K_{22}
            \end{array}\right)&  
            \footnotesize{\begin{array}{c}\arraycolsep=1mm
                  \left(\begin{array}{ccc}
                  0&1&0\\
                  1&0&0\\
                  0&0&0
                  \end{array}\right),\left(\begin{array}{ccc}
                  0&0&1\\
                  0&0&0\\
                  1&0&0
                  \end{array}\right),\left(\begin{array}{ccc}
                  0&0&0\\
                  0&0&1\\
                  0&1&0
                  \end{array}\right)\\\arraycolsep=1mm
                  \left(\begin{array}{ccc}
                  1&0&0\\
                  0&-1&0\\
                  0&0&0
                  \end{array}\right),
                  \left(\begin{array}{ccc}
                  0&0&0\\
                  0&1&0\\
                  0&0&-1
                  \end{array}\right)
            \end{array}}
            \\\\
            Oktupole&O_{ijk}=\left[\!\!\!\!\!\footnotesize{\begin{array}{c}\arraycolsep=1mm
            \left(\!\!\!\begin{array}{ccc}O_{111}&O_{112}&O_{113}\\O_{112}&O_{122}&O_{123}\\O_{113}&O_{123}&
            -O_{111}-O_{122}\end{array}\!\!\!\right)\\\\
            \left(\!\!\!\begin{array}{ccc}O_{112}&O_{122}&O_{123}\\O_{122}&O_{222}&O_{223}\\O_{123}&O_{223}&
            -O_{112}-O_{222}\end{array}\!\!\!\right)\\\\\arraycolsep=1mm
            \left(\!\!\!\begin{array}{ccc}O_{113}&O_{123}&-O_{111}-O_{122}\\O_{123}&O_{223}&-O_{112}-O_{222}\\
            -O_{122}-O_{122}&-O_{112}-O_{222}&-O_{113}-O_{223}\end{array}\!\!\!\right)
            \end{array}}\!\!\!\!\!\right] &
            \begin{array}{c}
           \text{7 tensor coefficients of}\\\\
          O_{111},\ O_{112},\ O_{122},\ O_{222},\\\\O_{113},\ O_{123},\ O_{223}
            \end{array}\\
    \end{array}
\label{MulSt}\end{equation}

Hence, tensors in the rightmost column enumerate the basis multipoles (and hence the basic states of a system of charges) for $n=0,1,2,3$. Instead that, one can enumerate the basis angular dependence tensors $\Omega_{i_1\ldots i_n}$ in the same way.

\begin{figure}[h!]
    \centering

     \includegraphics[width=1\textwidth,height=7cm,keepaspectratio]{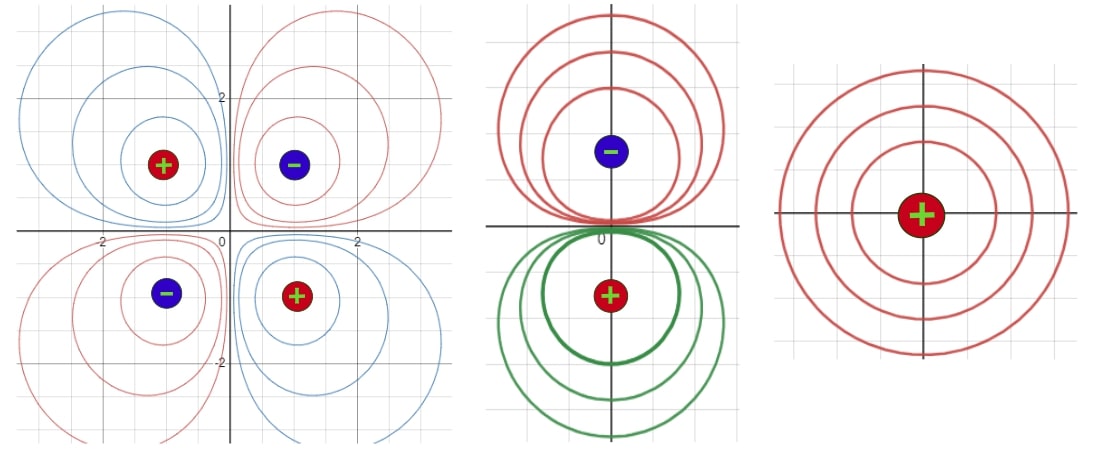}
    \caption{\footnotesize A quadrupole, a dipole, a monopole (a single charge).}
    \label{fig:MDK}

\end{figure}

\begin{figure}[h!]

        \centering
        \includegraphics[width=0.5\textwidth]{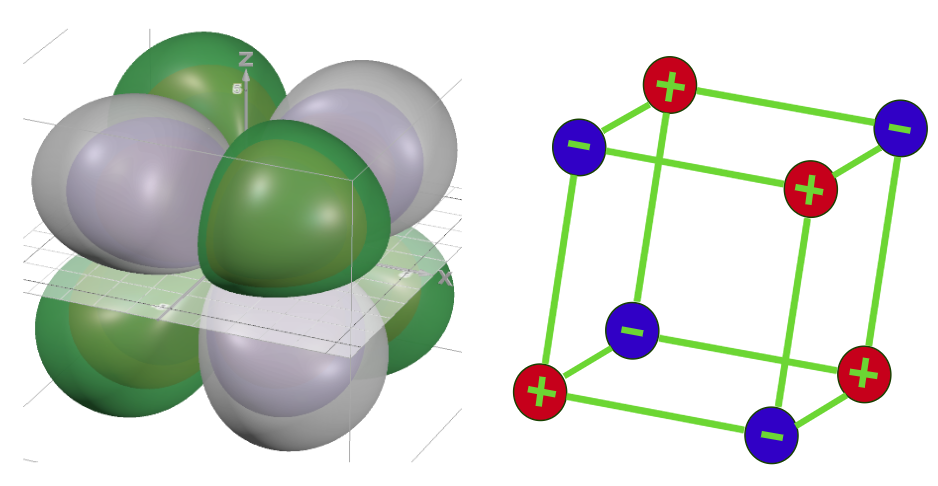}
    \caption{\footnotesize An oktupole.}
    \label{fig:Okt}

\end{figure}

\begin{figure}[h!]

    \centering
        \includegraphics[width=1\textwidth]{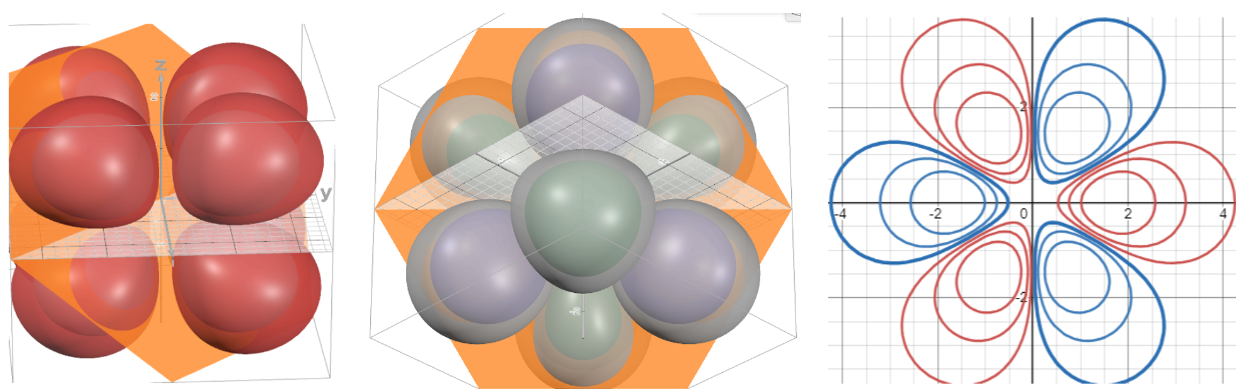}
    \caption{\footnotesize The section of octupole with the maximum symmetry.}
    \label{fig:Oktsec}

\end{figure}

\paragraph{Spherical harmonics.}
Given a particular multipole tensor, one can determine the angular dependence of the corresponding potential from (\ref{IntTay}). 
Examples of angular dependences for multipole potentials are given in Fig.\ref{fig:MDK},\ref{fig:Okt} and \ref{fig:Oktsec}. The shape of the Coulomb, dipole, quadrupole, and octupole potentials is surprisingly similar to how the $s$,$p$,$d$,$f$ orbitals are drawn in the chemistry textbooks. Apart from that, one can see the symmetry of the infinity, second, fourth, sixth order in the certain cross-sections. This makes us think that these potentials are described by trigonometric functions of the zeroth, single, double, and triple angle (in the corresponding planes). Indeed, one can make sure of that by substituting
$\vec n=(\sin\theta\cos\varphi,\sin\theta\sin\varphi,\cos\theta)$ and the particular form of the multipole tensors in (\ref{IntTay},\ref{MultTay}), e.g.,
\begin{equation}
\begin{array}{l}
Q=const\\\\
\Omega_jD_j^{(0)}=(n_x n_y n_z)\left(\footnotesize{\begin{array}{c}0\\0\\1\end{array}}\right)=\cos\theta,\\\\
\Omega_jD_j^{(1)}=(n_x n_y n_z)\left(\footnotesize{\begin{array}{c}1\\0\\0\end{array}}\right)=
\sin\theta\cos\varphi,\\\\
\Omega_jD_j^{(1')}=(n_x n_y n_z)\left(\footnotesize{\begin{array}{c}0\\1\\0\end{array}}\right)=
\sin\theta\sin\varphi\\\\
    \Omega_{ij}K^{(0)}_{ij}=
    \frac{1}{2}\mathrm{Tr\!}\left\{\!\!\left(\footnotesize{\begin{array}{ccc}
    3n_x^2-1&n_xn_y&n_xn_z\\n_xn_y&3n_y^2-1&n_yn_z\\n_xn_z&3n_yn_z&3n_z^2-1
    \end{array}}\right)\!\!\!\!
    \left(\footnotesize{\begin{array}{ccc}
                 -\frac{2}{3}&0&0\\
                 0&-\frac{2}{3}&0\\
                 0&0&\frac{4}{3}
            \end{array}}\right)\!\!\right\}=
            3\cos^2\theta-1=\frac{1}{2}(1+3\cos2\theta),\\\\
            \Omega_{ij}K^{(1)}_{ij}=
    \frac{1}{2}\mathrm{Tr\!}\left\{\!\!\left(\footnotesize{\begin{array}{ccc}
    3n_x^2-1&n_xn_y&n_xn_z\\n_xn_y&3n_y^2-1&n_yn_z\\n_xn_z&3n_yn_z&3n_z^2-1
    \end{array}}\right)\!\!\!\!
    \left(\footnotesize{\begin{array}{ccc}
                 0&0&0\\
                 0&0&1\\
                 0&1&0
            \end{array}}\right)\!\!\right\}=
            \sin\theta\cos\theta\sin\varphi=\frac{1}{2}\sin 2\theta\sin\varphi,\\\\
    \Omega_{ij}K^{(2)}_{ij}=
    \frac{1}{2}\mathrm{Tr\!}\left\{\!\!\left(\footnotesize{\begin{array}{ccc}
    3n_x^2-1&n_xn_y&n_xn_z\\n_xn_y&3n_y^2-1&n_yn_z\\n_xn_z&3n_yn_z&3n_z^2-1
    \end{array}}\right)\!\!\!\!
    \left(\footnotesize{\begin{array}{ccc}
                 0&1&0\\
                 1&0&0\\
                 0&0&0
            \end{array}}\right)\!\!\right\}=
            \sin^2\theta\sin\varphi\cos\varphi=\frac{1}{4}(1-\cos2\theta)\sin 2\varphi,
            \\\\
            \ldots
\end{array}\label{MultY}
\end{equation}
In the right part of (\ref{MultY}), one can recognise  real spherical harmonics $Y_{0,0}$,
$Y_{1,0}$, $Y_{1,1}$, $Y_{1,1'}$, $Y_{2,0}$, $Y_{2,1}$, $Y_{2,2}$, $\ldots$
(up to a normalisation) \cite{Jack,Vil}. These are the graphs of these functions that are drawn as ``shapes of the atom orbitals''.
This is not a coincidence, as we will see below.

\subsection{Multipoles and atom orbitals \label{sec:mulorb}}
\paragraph{Multipole ``quantum numbers'' $l$ and $m$.}
Generally, each real spherical harmonics has the two indices. Namely, one writes $Y_{l,m}(\theta,\varphi)$, where $l$ takes arbitrary nonnegative integer values, and $m$ can be equal to $0$, to an integer from $1$ to $l$, or to a ``integer with a prime'' from $1$ to $l$, instead of which one can formally use negative integers. Then $m$ takes values from $-l$ to $l$. Moreover, one can write $Y_{l,m}(\theta,\varphi)$ so that $l$ and $|m|$ are factors of $\theta$ and $\varphi$ in arguments of the trigonometric functions, and hence these numbers define the symmetry orders of a spherical harmonics in different planes. Note also that the number $l$ does not depend on the coordinate system (it determines the number of spherical functions of the given order), while the number $m$ is defined w.r.t. the selected $z$-axis of the spherical coordinates.  

Recall that in our case the spherical functions were obtained as the angular dependences of certain multipoles, and that there are $2l+1$ spherical functions of the order $l$ by definition.  Moreover, as we partially demonstrated in (\ref{MultY}), 
\begin{center}
\begin{tabular}{cp{15cm}}
$\bullet$ &One can take $2l+1$ basis multipoles or order $l$, $\left\{M^{(m)}_{i_1\ldots i_l}\big|m=-l,\ldots,l\right\}$, so that the angular dependences of the potentials are given by the spherical functions $Y_{l,m}(\theta,\varphi)$. 
\end{tabular}
\end{center}

Because the contributions of each order to the potential are scalars, and due to the transformation law of multipole tensors, one can see that the spherical harmonics turn under rotations into linear combinations of each other. Hence, they by definition span the representation space of the rotation group \cite{Vil}.

\paragraph{Quantum numbers in the atom model.}
Thus we see that
\begin{center}
\begin{tabular}{cp{15cm}}
$\bullet$ &
A far observable can relate to a system of electrostatic charges its space of states, where a basis is enumerated by ``quantum numbers'', namely, by integers $l$ and $m$ with
$ 0\le l$ and $-l\le m\le l$. 
\end{tabular}
\end{center}
Physically state of the system can be described twofold, either as an expansion of the charge density over the multipole tensors, or as expansion of the angular part of the potential over the spherical functions. If a state is \textit{pure}, i.e., it coincides with one of the selected basis states, then it is determined by \textit{quantum numbers} $l$ and $m$, which are either indices of the spherical function, or rank and number of the multipole tensor. Similar quantum numbers enter the description of an atom state. But in case of an atom it is \textit{principally} impossible to ``go inside'' and to know the exact position of each charged particle. One has to be content with the data about the system that are available for a far observable.

In the atom model \cite{LL3}, a (sub)orbital is given by a triple of quantum numbers, $(n,l,m,)$ where $l,m$ are so called orbital and magnetic quantum numbers, and $n$ is a principle quantum number or number of \textit{energy level}. Instead $n$ one can take the radial quantum number $n_r$. In our model the analogy of $n_r$ would be the power in the law of descending of the potential with the distance to the far observable, by tradition minus 1. But in our model $n_r=l$, while $n_r$ is an independent quantum number in atom model. This is due to the essential physical difference between these models, which we discuss below.

\subsection{Radiation and atom model\label{sec:multrad}}

\paragraph{Angular ``quantum numbers'' from Laplace equation.}
Up to now, we talked about the systems of electrostatic charges. But atom model arose as an attempt to describe atom \textit{radiation}. In classical ``raisins'' model the charged particles that constitute the atom oscillate around their positions and radiate the electromagnetic waves. Despite that generally it is much more difficult to describe a radiation of electro-magnetic waves than potential of system of static charges, the both problems are rather similar for a stationary radiation and for a far observable \cite{Jack}.

Indeed, one can derive the expansion of potential of a far system of charges over the spherical harmonics just from solution of the Laplace equation, in the limiting case, when all charges are placed near the origin,
\begin{equation}
    \Delta\varphi(\vec r)=0,\qquad r\ne 0. \label{Lap}
\end{equation}
The Laplace operator can we written down in spherical coordinates,
\begin{equation}
    \Delta=\Delta_r+\frac{1}{r^2}\Delta_{\theta,\varphi},\qquad
    \Delta_r=\frac{1}{r^2}\partial_r r^2\partial_r,\qquad
    \Delta_{\theta,\varphi}=\Big(\frac{1}{\sin\theta}\partial_{\theta}\sin\theta\partial_{\theta}+\partial^2_{\varphi\varphi}\Big).\label{Lapsph}
\end{equation}
This allows one to search for solutions of the Laplace equation by separating the variables, i.e., as a linear combination basis functions of the form $\phi(\vec r)=\chi(r)\Theta(\theta,\varphi)$. Then it follows from (\ref{Lap},\ref{Lapsph}) that the radial and angular factors must be \textit{eigenfunctions} of the corresponding addends in the operator $r^2\Delta$, 
\begin{equation}
    \Delta_r\chi(r)=\lambda \chi(r),\qquad
    \Delta_{\theta,\varphi}\Theta(\theta,\varphi)=-\lambda\Theta(\theta,\varphi),\qquad
    \lambda=const,
\end{equation}
where $\chi(r)$ tends to zero at the infinity (in order to the potential of the system of charges tended to zero far from it), and $\Theta(\theta,\varphi)$ is continuous on the unit sphere. But we already know from the multipole expansion (see sec.\ref{sec:symmult}), that the function 
$\phi^{(l,m)}(\vec r)=\frac{1}{r^{l+1}}Y_{l,m}(\theta,\varphi)$ for all positive $l$ are solutions to $(\ref{Lap})$ and have the desired form. Indeed, one can check intermediately that
\begin{equation}
    \Delta_r\frac{1}{r^{l+1}}=l(l+1)\frac{1}{r^l},\qquad
    \Delta_{\theta,\varphi}Y_{l,m}(\theta,\varphi)=-l(l+1)Y_{l,m}(\theta,\varphi),\qquad 
    \partial^2_{\varphi\varphi}Y_{l,m}=-m^2Y_{l,m}(\theta,\varphi),\label{LapY}
\end{equation}
where the third inequality allows one to enumerate functions $Y_{l,m}$ and to find their explicit form. It remains to check that all solutions of the original problem can be obtained as linear combinations of the above basis functions, and it indeed can be shown\footnote{The clue role here is played by the scalar product that can be introduced on the angular functions so that the basis functions $Y_{l,m}$ are orthonormal \cite{Vil}.}.

Hence,
\begin{center}
\begin{tabular}{cp{15cm}}
$\bullet$ & One can find a basis on the space of the solutions of the Laplace equation (the main electrostatic equation in the empty space), where each vector of the basis is given by the ``quantum numbers'', integers $l$ and $m$ such that 
$ 0\le l$ and $-l\le m\le l$. 
\end{tabular}
\end{center}

\paragraph{Classical radiation model and the radial ``quantum number''.}
On the other hand, the generalised potential $\Phi$ of the monochromatic electromagnetic wave of the frequency $\omega$ is described by Helmholz equation,
\begin{equation}
    \Delta\Phi=\omega^2\Phi,\label{Helm}
\end{equation}
where one can perform a separation of variables similar to (\ref{LapY}). I.e., one can find a basis in the solutions in form of the product of the spherical harmonics $Y_{l,m}(\theta,\varphi)$ and the radial function  $\chi(r)$, where the latter one satisfies
\begin{equation}
    \left(\Delta_r-\frac{l(l+1)}{r^2}-\omega^2\right)\chi=0.\label{Helmr}
\end{equation}
Thus $\chi$ depends not just on $l$, but on $\omega$ as well. Worse than that, the differential equation on 
$\chi$ now contains the coefficient that is explicitly dependent on $r$. Suppose that we are interested in the solution that tends to zero at the infinity, which could simulate a radiation \textit{far from} the system. Analysis shows \cite{LL3}, that there is such a solution only when $\omega$ runs over a certain, discrete set of values $\{\omega_n\}$. Hence, the basis in all radial functions that tend to zero at the infinity, $\chi(r,\omega_n,l)\equiv\chi_{n,l}(r)$, is labelled by one more quantum number $n$. In turn, basis in the entire space of the solutions of our interest is given by the functions $Y_{l,m}(\theta,\varphi)\chi_{n,l}(r)$, which (unlike the electrostatic problem) are enumerated by three independent ``quantum numbers'' $n$,$l$,$m$. Hence,
\begin{center}
\begin{tabular}{cp{15cm}}
$\bullet$ & One can find the basis on the space of solutions to the Helmholz equation (which describes a stationary monochromatic radiation), where each vector is labelled by the ``quantum numbers'' $n$,$l$,$m$.
\end{tabular}
\end{center}

\paragraph{Bohr atom and Schr\"odinger atom} 
Niels Bohr answered to the problem of an atom instability by a radical suggestion. He suggested to substitute an electron orbit in the usual sense by a quantity that is a function on the orbit and describes a state of an electron in the atom. Because a classical orbit is closed, the function must be periodic (i.e., it must describe a standing wave on the orbit).
Based on the symmetry of the problem, Bohr imagined himself circular orbits with linear combinations of the trigonometric functions $\sin(m\varphi)$, $\cos(m\varphi)$ (where $m$ is a nonnegative integer). Bohr postulated that the same is true for a general stated of an electron on an orbit, i.e., it can be represented as linear combination of \textit{pure} states, which are described by functions $\sin(m\varphi)$, $\cos(m\varphi)$. I.e., each state is described by an integer $m$, or by the number of  ``electron waves'' along the orbit, up to a phase shift, which is unobservable in the Bohr model. Hence the famous Bohr quantisation rule arises.

Erwin Schr\"odinger came further. He noticed that an electron in the field of nucleus generally possesses a spherical, not circular or cylindrical symmetry. And that there are no reasons to think that an unobservable ``electron orbit'' is a circle or even a line. Namely, Schr\"odinger suggested to describe an electron state by a quantity called \textit{a wave function} on the entire 3d space. Because the nuclear charges are localised near the origin, the problem for the electron possesses the symmetry w.r.t. general rotations around the origin. Hence, if one selects some subspaces in the space, they must be not \textit{circular} orbits, but \textit{spherical} orbitals. A sphere with the center in the origin turns into itself under general rotations w.r.t. the origin. Moreover, each point of the sphere can be turned into each point of the sphere by a rotation. Thus one says that a sphere is an irreducible invariant space of the rotation group w.r.t. the center of the sphere.

In analogy to the pure states on the Bohr circular orbit, one can introduce pure states on the Schr\"odinger spherical orbit. But the basis functions now are not trigonometric functions $\sin(m\varphi)$, $\cos(m\varphi)$, but spherical functions $Y_{l,m}(\theta,\varphi)$. I.e. the corresponding states are described already by the two quantum numbers, $l$ and $m$.

More rigorously, Schr\"odinger introduced a function $\psi(r,\theta,\varphi)$ and an equation on it, which describes an electron state in the atom. If an electron is not subjected to an impact explicitly dependent on the time, then the equation (similarly to the Laplas and Helmholz equations) has a solution in the form $\psi(r,\theta,\varphi)=Y_{l,m}(\theta,\varphi)\chi(r)$, where $\chi(r)$ satisfies a (\ref{Helmr})-like equation, up to the shift $l(l+1) \to l(l+1)+\kappa$ and the change $\omega^2$ for the number $\lambda$ with other physical sense. 
in (\ref{Helm}). The procedure of solution and the way of enumerating the solution are then completely similar. Hence,
\begin{center}
\begin{tabular}{cp{15cm}}
$\bullet$ & On the space of stationary solutions of stationary Schr\"odinger equation (which describes an electron state in the atom without a time-dependent impact), one can find a basis where each vector is labelled by the quantum numbers $n$,$l$,$m$.
\end{tabular}
\end{center}
\paragraph{Physics of quantum numbers.}
Now, having an idea about the atomic model, let's return to manifestations of the quantum numbers in the physical world.

Let's start from the Mendeleev table. Its modern interpretation is that the elements differ from each other both by a mass and a charge, and by a set of electrons in different states. Moreover, each electron stated is labelled by the quantum numbers $n,l,m$ (the triple for the electron that presents for the element and is absent for the previous element is given in the table near the element). Indirect confirmation of this model follow from physical and chemical properties of the elements, which change in a certain way depending on the set of electron states.

Atom radiation allows one to research atom in electron more accurately. According to the atom model, an electron in a state with quantum numbers $n,l,m$ has the certain energy $\mathcal{E}_{n,m,l}$. An electron transition into a state with quantum numbers $n',l',m'$ and with the smaller energy accompanies radiation of a quantum of light with the energy 
$\Delta \mathcal{E}=\mathcal{E}_{n,m,l}-\mathcal{E}_{n',l',m'}$ and with the corresponding frequency 
$\omega=\frac{\Delta\mathcal{E}}{\hbar}$ (where $\hbar$ is a Plank constant). The frequency is an observable quantity (the observation consists in passing a light through a spectroscope and obtaining well known pictures of atom spectra as a result). In this sense, spectral lines of the atom ``enumerates'' pairs of electron states and contain information about the states.

Moreover, the lines change in a certain way under various impacts on the atom. For example, the behaviour of spectral lines in a strong magnetic field allows one to conclude the energy  $\mathcal{E}_{n,m,l}$ in magnetic field  
$\vec H$ directed along $z$ (used to define $m$, see sec.\ref{sec:mulorb}) changes by $\delta \mathcal{E}\sim  \hbar m H$. This an a number of others observations allows one to relate the quantum number $m$ with the projection of the electron magnetic moment in the given state to the $z$ axis from the definition of $m$ \cite{LL3}. Hence $m$ is called a \textit{magnetic} \textit{quantum} \textit{number}. On the other hand, classical \textit{hyromagnetic analogy} \cite{LL2}, which considers an electron as a charged particle rotating around the nucleus, relates the electron magnetic moment with the its angular momentum. Hence,
\begin{center}
\begin{tabular}{cp{15cm}}
    $\bullet$ &The behaviour of the spectral lines allows one to relate the quantum number $m$ with the projection of the magnetic moment of the electron in the given state and with the projection of the angular momentum of the electron as if it was a classical particle rotating around the nucleus.
\end{tabular}
\end{center}

\subsection{What is the angular momentum operator?\label{sec:Lop}}
\paragraph{Quantum numbers and operators.}
Now we see that physics of the quantum number $m$ allows one to relate it to the projection of the electron angular momentum on the $z$ axis, with our definition of the $m$. On the other hand, solving the Schr\"odinger equation in analogy with (\ref{LapY}) and (\ref{Helm}), one gets the square of the quantum number $m$ as an eigenvalue of the operator 
$-\partial^2_{\varphi\varphi}$. It is convenient to perform here a trick, which is often used in description of oscillations and waves. Consider the complex operator $i\partial_{\varphi}$ (the ``square root'' of the operator $-\partial^2_{\varphi\varphi}$), and take in the solution space of (\ref{LapY}) 
a new basis $\tilde{Y}_{l,m}$ by substituting each pair of the real functions
$Y_{l,m}=X_{l,m}(\theta)\cos (m\varphi)$, $Y_{l,m'}=X_{l,m}(\theta)\sin (m\varphi)$ with the pair of the complex functions  
\begin{equation}
\tilde{Y}_{l,\pm m}=Y_{l,m}\pm iY_{l,m'}= X_{l,m}(\theta)e^{\pm im\varphi}, \qquad
Y_{l,m}=\Re\,\tilde{Y}_{l,\pm m},\ Y_{l,m'}=\pm\Im\,\tilde{Y}_{l,\pm m}, 
\qquad m>0.  
\end{equation} 
One then can introduce the complex potential 
$\tilde\phi^{l,m}=-r^{-l-1}\tilde{Y}_{l,m}(\varphi,\theta)$, which describes a \textit{pair} of physical potentials $\Re\,\tilde{\phi}^{l,m}$, $\Im\,\tilde{\phi}^{l,m}$.
Then the operator $-i\partial_{\varphi}$, which in Cartesian coordinates equals $i(y\partial_x-x\partial_y)\equiv \hat L_z$, satisfies
\begin{equation}
\hat L_z\tilde{\phi}^{l,m}=m\tilde{\phi}^{l,m},\qquad\Leftrightarrow\qquad
\left\{\begin{array}{rcl}
\partial_{\varphi}\phi^{(l,m)}&=&-m\phi^{(l,m)}\\
\partial_{\varphi}&=&m\phi^{(l,m')}
\end{array}\right.
\end{equation}
Due to the physical sense of the quantum number $m$ (see sec.\ref{sec:multrad}) operator $\hat L_z$ is called an \textit{angular momentum projection operator}\footnote{Standard definition of the angular momentum operator differs by the factor of the Plank constant $\hbar$. Then in the corresponding experiments, the eigenvalues of the angular momentum operator are conserved in the sum with the classical angular momentum without extra factors (see sec.\ref{sec:QclL}.)} on the $z$ axis. 
One can introduce the operators $\hat L_x$ and $\hat L_y$ in a similar way. The thus obtained triple of the operators can be considered as the three components of the vector operator
\begin{equation}
\hat L\equiv-i[\vec r\times \nabla]=
\Big(i(z\partial_y-y\partial_z),i(x\partial_z-z\partial_x), i(y\partial_x-x\partial_y)\Big)
\equiv\Big(\hat L_x, \hat L_y, \hat L_z\Big),
\end{equation}
which is called \textit{angular momentum operator}. Hence,
\begin{center}
\begin{tabular}{cp{15cm}}
    $\bullet$& The operator, whose eigenvalue is the magnetic quantum number $m$, is called an operator of angular momentum projection to the $z$ axis used to define the quantum number $m$.\\
    $\bullet$& This operator is a component of the vector angular momentum operator.
\end{tabular}
\end{center}
The case of the orbital quantum number $l$ is more difficult. On the one hand, $l$ is the maximum absolute value of $m$. On the other hand the number $l$ itself is not an eigenvalue of an operator with a clear physical sense. Yet the number $l(l+1)$ is the eigenvalue of the angular part of the Laplace operator
$\Delta_{\theta,\varphi}$. To relate this quantity with the angular momentum operator, we note that
\begin{eqnarray}
\boxed{r^2\Delta}=(x^2+y^2+z^2)(\partial^2_{xx}+\partial^2_{yy}+\partial^2_{zz})=
(x\partial_x+y\partial_y+z\partial_z)^2-(x\partial_x+y\partial_y+z\partial_z)+
\nonumber\\
+(x\partial_y-y\partial_x)^2+(z\partial_x-x\partial_z)^2+(y\partial_z-z\partial_y)^2+
2(x\partial_x+y\partial_y+z\partial_z)=\nonumber\\=
(\vec r\cdot \nabla)^2+\vec r\cdot \nabla+[\vec r\times\nabla]^2\ \boxed{=\partial_r(r^2\partial_r)+\hat L^2},
\nonumber\\
\label{LapRadAng}
\end{eqnarray}
Comparing (\ref{LapRadAng}) with (\ref{Lapsph}), we get that 
$\hat L^2=\Delta_{\theta,\varphi}$. Hence,
\begin{center}
\begin{tabular}{cp{15cm}}
    $\bullet$& The orbital quantum number $l$ gives the maximum absolute value of the magnetic quantum number $m$ and the eigenvalue of the square momentum operator that equals $l(l+1)$.
\end{tabular}
\end{center}

\paragraph{``$\mathbf{l^2=l(l+1)}$''.}
Here one can see one of the quantum mechanical paradoxes, ``the square of the angular momentum does not equal the the square of the maximum projection of the angular momentum''.
The case here is that the eigenvalue of the vector operator component is not the same as an ordinary vector component. If the projection of the vector $\vec v$ on a give $z$-axis takes its maximum value $v_z=v$, then $v_x=v_y=0$ and $\vec v^2=v^2$. But if a component of the vector operator $\hat L$ has the maximum eigenvalue $l$ for a vector (i.e., for an element of a linear space) $\psi$, then this vector must not have the eigenvalue 0 for the other two components. Moreover, it must not be even an eigenvector for them!

For example, consider a vector operator on the 2d linear space,
$\hat A=\left(\hat A_x, \hat A_y\right)$, where
$\hat A_x=\left(\begin{array}{cc}     1&0  \\     0&-1 \end{array}\right)$, and
$\hat A_y=\left(\begin{array}{cc}     0&1  \\    1&0 \end{array}\right)$.
The operator $\hat A_x$ has the two eigenvectors, 
$\psi_+=\left(\begin{array}{c}     1  \\     0 \end{array}\right)$ and 
$\psi_-=\left(\begin{array}{c}     0  \\     -1 \end{array}\right)$, with the eigenvalues 1 and -1, $\hat A_x\psi_+=\psi_+$, $\hat A_x\psi_-=-\psi_-$. I.e., the maximum eigenvalues of the component $\hat A_x$ equals $1$ and is obtained for the eigenvector $\psi_+$. The operator $\hat A_y$ acts on this vector as $\hat A_y\psi_+=\psi_++\psi_-$, i.e., this vector is not an eigenvector for this operator.  Yet $\psi_+$ is the eigenvector on the operator $\hat A^2\equiv\hat A_x^2+A_y^2=
\left(\begin{array}{cc}     2&0  \\     0&2 \end{array}\right)$, but with the eigenvalue other than $1^2$: $\hat A^2\psi_+=2\psi_+$. 

\paragraph{Space rotation as a process of measurement.}
The angular momentum operator has a clear geometrical sense. The operator $\vec L$ is an operator of an infinitesimal rotation. Namely, a vector $\vec r$, when uniformly rotated around a unit vector $\vec n$ with the angular velocity $\omega$ changes as 
$\frac{d\vec r}{dt}=\omega\vec n\times r$. Hence for a small time $\delta t$, the vector rotates by the angle $\delta\varphi=\delta t \omega$ and thus changes by $\delta \vec r=\delta\varphi\,\vec n \times \vec r$. Then any scalar function of $\vec r$, i.e., potential of the system of charges, changes by
\begin{equation}
    \delta \phi(\vec r)=\delta \vec r\cdot\nabla\phi(\vec r)=\delta\varphi([\vec n \times \vec r]\cdot\nabla)\phi(\vec r)=
    \delta\varphi(\vec n \cdot [\vec r\times \nabla])\phi=\delta\varphi\,\vec n\cdot i\hat L\phi(\vec r).\label{momrot}
\end{equation}
On the other hand, the \textit{state space} of our system conserves under the rotations (see sec.\ref{sec:mulstate}).  In particular, that means that if the initial state of the system was a state of the selected basis, then the rotated state (as the vector of the same state space) is expanded into a linear combination of the basis states, 
\begin{equation}
\frac{\delta\phi^{(l,m)}}{\delta\varphi}=\boxed{\vec n\cdot i\hat L\phi^{(l,m)}=\sum_{\tilde{m}}C_{m,\tilde{m}}\phi^{(l,\tilde{m})}}.
\label{rotL}\end{equation}
In particular, if $m$ is defined w.r.t. $z$-axis,
\begin{equation}
\frac{\delta\phi^{(l,m)}}{\delta\varphi}=\boxed{i\hat L_z\phi^{(l,m)}=m
\phi^{(l,m)}}.
\label{rotLz}\end{equation}
In the language of sec.\ref{sec:RBLight}, a space rotation of the system is a variant of its evolution. Moreover, if the properties of the system do not depend on time, on can equally consider as an evolution parameter both the time and the rotation angle. In analogy with (\ref{transeq}), relations (\ref{rotL},\ref{rotLz}) then define the evolution paths on the state space of the system. We also discussed in sec.\ref{sec:RBLight}, that the evolution process can be considered as a measurement process of a quantity that is the same for all evolution paths. In case of general rotation (\ref{rotL}), such a quantity is $l$ as the number of evolution paths on each step. In case of rotation (\ref{rotLz}) around the $z$-axis w.r.t. which $m$ is defined, such a quantity is also $m$ as a coefficient of the only evolution path. 

Because a general \textit{small} rotation is given in (\ref{rotL}) as a liner combinations of the three rotations related to the angular momentum components, it is sufficient to consider the three rotations generated by the operators $i\hat L_x$, $i\hat L_y$, $i\hat L_z$ to measure the $l$.

Now we see one more interpretation of the angular momentum.
\begin{center}
\begin{tabular}{cp{15cm}}
    &For the $(l,m)$ state,\\
    $\bullet$&Operator of angular momentum projection $i\hat L_z$ generates an evolution of the state that is a measurement process of the quantum number $m$ defined w.r.t. the $z$ axis.\\
    
    $\bullet$& A linear combination of the three angular momentum components 
    $\hat L =\left(\hat L_x, \hat L_y, i\hat L_z\right)$ other than $\hat L_z$ generates an evolution of the state that is a measurement process of the quantum number $l$.
\end{tabular}
\end{center} 

\paragraph{What can be measured simultaneously?}
And what about measuring two projections of angular momentum of different axis? As we see from (\ref{rotLz}), a measurement of $m$ was possible for such state of the system, which turned into itself under the evolution generated by $\hat L_z$. The a simultaneous measurement of $m_1$ and $m_2$ defined w.r.t. different axis (say, $y$ and $z$) would be possible in such a state that turns into itself under rotations both around the $z$ and $y$ axis,
\begin{equation}
    \hat L_z\phi=m_1\phi,\ \ \hat L_y\phi=m_2\phi.\label{Rotyz}
\end{equation}
There is indeed such a state, namely it is the basis state $\phi^{(0,0)}$,
\begin{equation}
    \hat L_x\phi^{(0,0)}=\hat L_y\phi^{(0,0)}=\hat L_z\phi^{(0,0)}=0.
\end{equation}
Yet it is impossible to find a basis of states with property (\ref{Rotyz}) on the state space. Because the \textit{both} operators $\hat L_z$ and $\hat L_y$ would be given in such a basis by diagonal matrices, and then $\hat L_z\hat L_y=\hat L_y\hat L_z$ was satisfied. But $\hat L_z\hat L_y-\hat L_y\hat L_z=-i\hat L_x$ in any basis, as it follows from the definition of these operators.

However, our basis states by definition allow one to measure the quantum numbers $m$ and $l$ simultaneously. I.e., each basis state is an eigenstate both for the vectors $\hat L_z$ and $\hat L^2$. This is possible due to the relation $[\hat L^2,\hat L_z]=0$ (and similarly for $\hat L_x$ and $\hat L_y$), which can be checked independently.

Hence,
\begin{center}
\begin{tabular}{cp{15cm}}
    $\bullet$ &One can not measure simultaneously two angular momentum projections, i.e., two quantum numbers $m_1$ and $m_2$, defined w.r.t. different axis.\\
    
    $\bullet$ &One can measure simultaneously an angular momentum projection on any axis, i.e., the quantum number  $m$, defined w.r.t. any axis, and the quantum number $l$, related to the angular momentum square $l(l+1)$.
    \end{tabular}
\end{center}

\subsection{Quantum and classical angular momenta\label{sec:QclL}}
\paragraph{Angular momentum of a test particle.}
And where is in all this story a classical angular momentum? Let's return to the system of far charges.

Suppose that we in principle can know something about the system of charges only by its action on other systems that are near the observable (and thus far from the original system).
Consider a \textit{test charge} of magnitude $q$, i.e., a particle that moves under the action of system of fixed charges and is far from this system. The equations of motion can be presented in the form
\begin{equation}
    \cfrac{d \vec p}{d t}=-q\vec \nabla \phi\qquad\Rightarrow\qquad     
    \cfrac{d [\vec{r}\times\vec p]}{d t}=-q[\vec r \times \vec \nabla] \phi\qquad\Rightarrow\qquad
    \boxed{\frac{d\vec L}{dt}=-qi\hat L\phi}\ ,\label{momentop}
\end{equation}
where $\vec r$, $\vec p$, $\vec L$  -- are the radius vector, momentum and angular momentum of the test charge, and 
$\hat L$ is the already familiar angular momentum operator.
Hence,
\begin{center}
\begin{tabular}{cp{15cm}}
    $\bullet$& For a classical system of charges, the rate of change of the test particle angular momentum is given by the action of the angular momentum projection operator on the potential of the system. 
\end{tabular}
\end{center}
Let the system be in a state chosen as one of the basis states. Then
\begin{equation}
    \frac{dL_z}{dt}=i\hat L_z\tilde{\phi}^{(l,m)}=m\tilde{\phi}^{(l,m)}.\label{Lprob}
\end{equation}
I.e., if we can measure $\tilde{\phi}^{(l,m)}$ in a given point w.r.t. the infinity, then the ``quantum number'' $m$ is a measure of impact of the charge system on the angular momentum of the test charge. 

Recall that $\tilde{\phi}^{(l,m)}$ is a complex potential to a pair of physical ones. Actually (\ref{Lprob}) means that the operator $\hat L_z$ mixes these two potentials with each other (but not with other basis states), and that the mixing matrix depends on the only parameter $m$.

\paragraph{A flow of test particles and the operator of the squared angular momentum.}
Consider now an enough dense \textit{flow of test charges},  so that one can consider that in a space area (far from the original system) there is a test charge at each point. Then one can introduce the functions $\rho_q(\vec r)$, $\rho_m(\vec r)$, $\vec p(\vec r)$ that are the charge, mass and momentum densities of the test particles. There is a relation between these functions. Namely, consider a parallelepiped near a fixed point, whose edges are  parallel to the coordinate axis $x$, $y$, $z$ and equal to the small quantities $\delta x$, $\delta y$, $\delta z$. Then the total mass of the particles that pass through the pair of faces normal the the $x$ axis for the time $\delta t$ equal $p_x(x)\delta y\delta z\delta t$ and $p_x(x+\delta x)\delta y\delta z\delta t$. Hence, the mass  $\partial_x p_x(x)\delta V\delta t$, where $\delta V=\delta x\delta y \delta z$ is the volume of the parallelepiped, stays between the faces. Consider all three pairs of the parallel edges in a similar way, and obtain that the mass of the particles inside the parallelepiped overall changes by
$\delta m=\nabla \cdot \vec p\delta V\delta t$, so that
\begin{equation}
    \frac{d \rho}{dt}=\nabla\cdot \vec p\label{divp}.
\end{equation}
Herefrom, using the first of relations (\ref{momentop}), and supposing that the density of the test particles little changes in the volume of the parallelepiped, one gets
\begin{equation}
    \frac{d^2 \rho_m}{dt^2}=\nabla\cdot \rho_q\nabla \phi\approx \rho_q\Delta\phi.\label{LapPhi}
\end{equation}
Equation (\ref{divp}) allows one to give a physical sense to the Laplace operator in our system.
Yet also $\Delta\phi=4\pi\rho$, where $\rho$ is a density of charges, which is non-zero only near the origin. Hence in our case the both parts of (\ref{divp}) are identically zero. However, as we saw in (\ref{LapRadAng}), one can divide the Laplacian into two parts, radial and angular ones, and the angular part is proportional the square of the momentum operator $\hat L^2$.
Considering a flow of test particles as above, but for a volume element in the spherical coordinates, one can show that the operator
$\frac{1}{r^2}\partial_r(r^2\partial_r)$ gives exactly the radial component of the flux through a unit volume; hence the operator $\frac{1}{r^2}\hat L^2$ must give the remaining flux, along the coordinate sphere. One can make sure by straightforward computation of the angular part of the flux that the square momentum operator equals to the already obtained expression,
\begin{equation}
\hat L^2=\frac{1}{\sin\theta}\partial_{\theta}\sin\theta\partial_\theta+\partial^2_{\phi\phi},
\end{equation}
and thus indeed depends only on the angular coordinates.

\begin{center}
\begin{tabular}{cp{15cm}}
$\bullet$&The square momentum operator applied to potential of the original system equals to the acceleration of change of the density of test charges due to the angular (or the radial with the minus sign) part of the flux.    
\end{tabular}
\end{center}

\paragraph{Classical and quantum magnetic moments.}
The state space constructed in sec.\ref{sec:symmult} for a classical system of charges can be used to describe a much more general system with central symmetry \cite{LL3}. We partially discussed it in sec.\ref{sec:multrad}. In quantum mechanics, such description is not only applicable to many systems, but turns out to be the most complete one in the framework of quantum restrictions. Moreover, not only composite systems possess the described $(l,m)$ state space, but to elementary particles as well. The most surprising, this is applied to the true elementary particles, like electron, for which no internal structure is known at the moment. I.e., the $(l,m)$ states can describe both ``visible'' (at least in some sense) electron states in atom, and the ``internal'' states of the electron itself. Then in analogy with the physical sense of the numbers $(l,m)$ for composite systems, the particle is said to have an \textit{own torque} or a \textit{spin}. In this case, the same particle has always a constant quantum number $l$ (the ``spin magnitude''), while the quantum number $m$ (the ``spin projection'') can vary\footnote{Yet in case of composite particle there is a question, what to consider as different particles, and what to consider as different states of the same particles \cite{PDG}.}. There is an analogy with a light polarisation. Namely, the light has exactly two independent polarizations, but a particular light beam can take any of them, or (as usual) any mixture of them \cite{LL2}. By the way, a light polarisation is just related to the spin of the light quantum or photon in quantum physics\footnote{However, one needs slightly different model for the photon spin. This is because the photon moves with the light velocity and, according to the special relativity, can not rest in any frame \cite{FeyQED}. But our model of the $(l,m)$ states is related exactly to a rest particle.}.

In classical physics, the own torque of a system of charges is related to its magnetic moment. E.g., one can easily see this for a current frame. In a more general case it is one of the manifestations of the so called \textit{hyromagnetic analogy} \cite{LL2}. For quantum particles, one can usually observe exactly the magnetic moment, which on the grounds of the hyromagnetic analogy is related to the spin. The most intermediate observation of the magnetic moment of the electron (and other particles) provides Stern--Gerlach experiment. In this experiment, a beam of the quantum particles with the own magnetic moment placed in a inhomogeneous magnetic field divides into fractions with certain projection of the magnetic moment onto the mean projection of the field. This is exactly how a beam of classical magnetic moments would behave under the experimental conditions. The difference is that the projection of the magnetic moment of a quantum particle runs integer values $m$ from $-l$ to $l$, i.e., it quantises. Stern--Gerlach experiment and its relation to the ``internal state space'' of a quantum particles is presented in details in Vol.8 of the Feynman lectures on physics \cite{FLF8}. It is also explained there what means $l=\frac{1}{2}$ for the electron.

\paragraph{Quantum-classical conservation laws.}
It is even more surprising, that a quantum particle can ``pass'' its quantum momentum to a classical system as a classical angular momentum. I.e., 
\begin{center}
\begin{tabular}{cp{15cm}}
$\bullet$& If a system with a conserved angular momentum projection on a given $z$ axis, is composed of a classical subsystem with the angular momentum $\vec L$ and a quantum particle in the $(l,m)$ state, then the conserved quantity equals $L_z+\hbar m$. Moreover, each addend can vary at the expense of the other.
\end{tabular}
\end{center}
An important point is that the dimensionless integer $m$ comes with the factor of $\hbar$, which is the Plank constant. It has the dimension of angular momentum and relates here the quantum number $m$ with the angular momentum in physical units. The Plank constant is a fundamental constant that enters each relation between quantum and classical observables and is monstrously small. Hence in our case, it is hardly possible to notice a variance of the $L_z$ at the expanse of variance of the $m$. Yet it is possible. If a system contains enough many quantum particles, then very subtle experiments allow one to notice a transition of the quantum angular momentum to the classical one\cite{EdHHist}.

In 1908, O.Richardson supposed that a ferromagnetic sample will rotate when the magnetic field is turned on. This is because the magnetic moments of some ``small magnets'' that compose the sample line up along the magnetic field under magnetisation, and thus the sample gets a magnetic moment. But at the lack of the external torques, the classical hyromagnetic analogy predicts that the sum of mechanical angular momentum and magnetic moment with certain coefficient (the inverse quantity to which is called classical \textit{hyromagnetic ratio}) is conserved.
hyromagnetic analogy. Hence, the sample together with its magnetic moment must get the angular momentum of the corresponding magnitude and of the inverse sign. Hence, a ferromagnetic sample must start rotating under magnetisation. The inverse effect, which consists in magnetisation of the rotated sample, was predicted in 1909 by S.Barnett.

In 1915, A.Einstein and W. de Haas experimentally observed rotating of a sample under magnetisation (now the phenomenon is called \textit{Einstein--de Haas effect}). Even earlier, in 1914, S.Barnett observed magnetisation of a rotated sample (now the phenomenon is called \textit{Barnett effect}). His results were quantitatively specified in 1918 by J.Steward.

A similar phenomenon for a light quantum (photon), i.e., a transition of the own torque to the mechanical magnetic moment, was predicted in 1898 by A.Sadowsky. Now we can say that a circular polarized light consists of photons with certain spin projection on the light ray, and such light at normal incidence must rotate a plate. Similar to the case of electron spin, an affect of the quantum quantity on the classical one is monstrously small. But precision experiments performed by R.Bethe in 1936 made it possible to observe the effect for the first time  \cite{SadNew}.

\section{From closed orbits to hidden symmetries\label{sec:hidsymm}}
By now we considered enough visual symmetries, such as translations and 3d rotations. In this section, we consider a much less obvious and intuitive, but rather curious symmetry of the problem of motion in the $k/r$ potential. Because this potential describes both Coulomb and Newtonian interaction, we will consider in parallel the classical problem of planet motion in the the Newtonian gravitating  centre and the quantum problem of electron states in the Coulomb field of the nucleus\footnote{In this section, we mean by an atom a hydrogen atom, whose nucleus consists of the single proton (which can be in this case considered as a point-like charge), so that the field of the nucleus is indeed a Coulomb field.}. We will treat them classical and quantum versions of the same phenomenon.  

The main object of our interest here is the Laplace--Runge--Lenz (LRL) vector and its quantum analogue.
The story about symmetries of the $k/r$ potential is revealed more profoundly in \cite{HydSymm}. 

\subsection{Perihelion conservation law}
\paragraph{Conservation laws in Kepler problem.}
The Kepler problem is a particular case of a two-body problem, where the potential energy decreases proportionally to the distance to the appealing centre. Then the interaction force of a material point with the centre is directed to the centre and proportional to the inverse square of the distance to the centre,
\begin{equation} \label{field}
U(r) = -\frac{k}{r},\qquad \vec{F} = -\vec{\nabla}{U} =  -\frac{k}{r^3}\vec{r}
\end{equation}
Consider a motion of the material point in this problem and find the conserving quantities. One shows in theoretical mechanics that each conserving quantity is related to a symmetry of the system \cite{Arn}. In our case, the system is symmetric w.r.t. a time shift, and this is associated with the energy conservation $E = \frac{m\dot{\vec{r}}^2}{2} - \frac{k}{r}$. In addition, the interaction in the problem is central, i.e., there is the full rotation  rotation symmetry, which is associated with conservation of the angular momentum $\vec{L} = [\vec{r}, \vec{p}]$. The both conservation laws can be also checked straighforwardly. Conservation of the angular momentum implies that the motion is planar. Indeed, the vectors $\vec{r}$ and $\vec{p}$ always lie in the same plane normal to the constant vector $\vec{L}$,
\begin{equation} \label{plane}
(\vec{r}, \vec{L}) = 0, \qquad (\vec{p}, \vec{L}) = 0.
\end{equation}
One more conserving quantity is the so called LRL vector,
\begin{equation} \label{A}
\vec A = [\vec p, \vec L] - mk\vec n,
\end{equation}
where $\vec n = \frac{\vec r}{|\vec r|}$.
We prove this fact using the Newton second law,
\begin{equation} \label{Newton}
\frac{d\vec p}{dt} = -\frac{k}{r^3}\vec{r}.
\end{equation}
Consider a motion of the unit vector $\vec n=\vec r/r$, which rotates with the angular veclocity $\Omega=\vec L/r^2$,
\begin{equation} \label{rhat}
\frac{d\vec n}{dt} =[\vec \Omega, \vec n] =-\frac{[\vec r, \vec L]}{mr^3}
\end{equation}
Then
\begin{equation} \label{cons}
\frac{d\vec A}{dt} = \left[\frac{d\vec p}{dt}, \vec L\right] + \left[\vec p, \frac{d\vec L}{dt}\right] - mk\frac{d\vec n}{dt} = 0
\end{equation}
where we also used the conservation of the angular momentum.

Note that $(\vec A, \vec L) = 0$, i.e., the RLR vector is always orthogonal to the angular momentum vector.

\paragraph{Planets moves in ellipses.}
Because the motion is planar (as we saw above), we can consider the particle trajectory in polar coordinates ($\rho, \phi$). We put the origin at the appealing centre, i.e., $\rho = |\vec r|$, and we measure the angle from the LRL vector. Consider the scalar product $(\vec A, \vec r)$:

\begin{gather*}
(\vec A, \vec r) = A\rho\cos\phi\\
(\vec A, \vec r) = (\vec r, [\vec p, \vec L]) - mkr = L^2 - mk\rho
\end{gather*}
\begin{equation} \label{tr}
    \rho = \dfrac{\dfrac{L^2}{mk}}{1+\dfrac{A}{mk}\cos\phi}
\end{equation}

The trajectory is a conic, or a second order curve \cite{Arn}. Consider the eccentricity $e = \frac{A}{mk}$,
\begin{equation} \label{e}
    e = \frac{A}{mk} = \sqrt{1 - \left(\frac{p_0L}{mk}\right)^2}
\end{equation}
This expression is checked by straightforward substitution of (\ref{A}) into (\ref{p0}).
Then  the particle moves in ellipse if $E > 0$, in hyperbola if $E < 0$, and in parabola if $E = 0$.
We can also express the major and minor axis (in case of ellipse) or the real and imaginary axes (in case of hyperbola),
\begin{gather} \label{axis}
    a = \dfrac{l}{1 - e^2} = \dfrac{mk}{p_0^2}\\
    b = \dfrac{l}{\sqrt{1 - e^2}} = \dfrac{L}{p_0},
\end{gather} 
where $a$ is the major or real axis, $b$ is the minor of imaginary axis, and $l = \dfrac{L^2}{mk}$ is a focal parameter.

\begin{figure}[h!]
    \centering

     \includegraphics[width=1\textwidth,height=7cm,keepaspectratio]{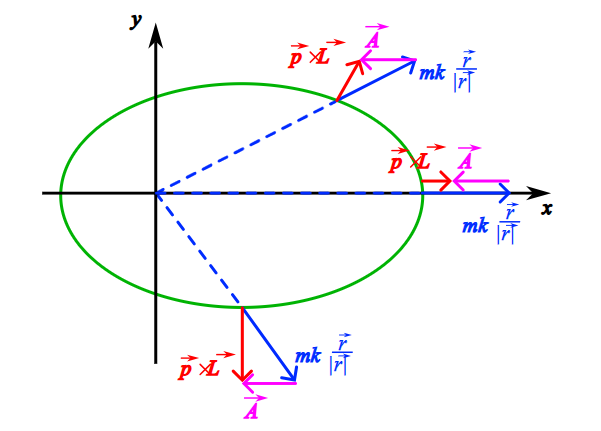}
    \caption{\footnotesize A finite trajectory in Kepler problem. 
    The figure also shows how the LRL vector is divided into two addends according to (\ref{A}).}
    \label{fig:ellipse}

\end{figure}

\paragraph{The end of the vector moves in a circle.}  Using the conservation of the LRL vector, one can obtain the particle trajectory in the momentum space (momentum hodograph). Because $\vec p \perp \vec L$ and $\vec A \perp \vec L$, the following identities are true,
$$[\vec p, \vec L]^2 = |\vec{p}|^2|\vec{L}|^2,$$
$$[\vec A, \vec L]^2 = |\vec{A}|^2|\vec{L}|^2.$$
Now express the $mk\vec n$ from (\ref{A}) and divide over the $L^2$,
\begin{gather*}
\frac{(mk\vec n)^2}{L^2} = \frac{([\vec p, \vec L] - \vec A)^2}{L^2} = \\
= p^2 - 2(\vec A, [\vec p, \vec L]) + \frac{[\vec A, \vec L]^2}{L^4} = \\
= \left(\vec p - \frac{[\vec A, \vec L]}{L^2}\right)^2,
\end{gather*}
where the symmetry of the mixed vector product w.r.t. a cyclic permutation was used.

\begin{figure}[h!]
    \centering

     \includegraphics[width=1\textwidth,height=7cm,keepaspectratio]{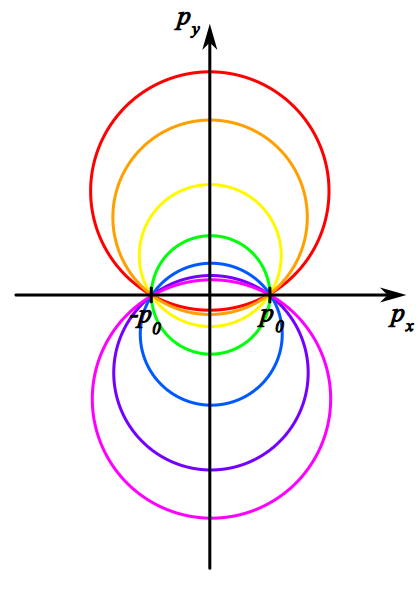}
    \caption{\footnotesize Momentum hodograph. The figure shows hodographs for the same positive energy and different values of $A$ and $L$.
    }
    \label{fig:godograph}

\end{figure}

Finally, we obtain the equation of a circle,
\begin{equation} \label{hod}
    \left(\vec p - \frac{[\vec A, \vec L]}{L^2}\right)^2 = \frac{(mk)^2}{L^2}
\end{equation}
If we direct the $p_z$ axis along the $\vec L$ and we direct $p_x$ axis along the $\vec A$, then we obtain a family of circles (Fig. \ref{fig:godograph})
$$
    p_x^2 + \left(p_y - \frac{A}{L}\right)^2 = \frac{(mk)^2}{L^2}.
$$
I.e., all three types of trajectories in coordinate space correspond to a circle in the momentum space. The finitness criterion is here as follows (it is rather visual, but we derive it below). Let's see whether the circle intersects the $p_x$ axis (in case of two intersections $p_x\ne 0$, $p_y=0$ we have the two turning points of the finite trajectory). Consider the quantity $p_0^2 = \frac{(mk)^2 - A^2}{L^2}$. Now we find the conditions under which $p_0^2$ is positive, i.e., there is an intersection; there is no intersection in the opposite case.
\begin{gather*}
p_y = 0,\quad p_x = p_0,\quad p = p_0\\
A = A_x = Lp_y - mk\frac{x}{r} = -mk\frac{x}{r}\\
L = xp_y - yp_x = yp_0\\
p_0^2 = \frac{(mk)^2 - A^2}{L^2} = (mk)^2\frac{(r)^2 - (x)^2}{(ryp_0)^2} = \left(\frac{mk}{rp_0}\right)^2\\
\frac{k}{r} = \frac{p_0^2}{m}
\end{gather*}
Now consider the total energy
\begin{gather*}
E = \frac{p_0^2}{2m} - \frac{k}{r} = -\frac{p_0^2}{2m},
\end{gather*}
then
\begin{equation} \label{p0}
    p_0^2 = -2mE.
\end{equation}
I.e., if $E < 0$, the hodograph intersects the $p_x$ axis at the points $\pm p_0 = \pm\sqrt{2m|E|}$, and the corresponds to the finite motion.\\
If $E > 0$, the hodograph does not intersect the $p_x$ axis, and the motion is infinite.\\
If $E = 0$, the hodograph is tangent to the $p_x$ axis, and the motion is also infinite.

\paragraph{Scaling invariance.}
The Kepler problems has one more symmetry property, which is conserved in a more general case of motion in the $\frac{a}{r^{\alpha}}$ potential. Namely, the equation of motion
$m\frac{d^2\vec r}{dt}=-\vec\nabla \frac{a}{r^{\alpha}}$ do not change under the simultaneous transformation of the coordinates \cite{Arn} $\vec r\to\lambda^2 r$ and
$t \to \lambda^{\alpha+2}t$. In case $\alpha=1$ the transformation takes the form $\vec r\to\lambda^2 r$ and 
$t \to \lambda^3t$, and here from the Kepler third law follows. Moreover, this transformation preserves the combination $EL^2$ together with $A^2$ (as follows from the explicit expression for the LRL vector). In addition, the transformation preserves the direction of the vector in the space, so that the very vector $\vec  A$ is also invariant.

\subsection{A hidden classical symmetry\label{sec:clRLR}}
\paragraph{A trick with classical symmetries.}
In sec.\ref{sec:RBLight}, we introduced the evolution operator of a quantum system in time, in particular of evolution for a small time $\tau$ (see sec.\ref{sec:evoexp}).
For a classical system with the conserved energy $E=\cfrac{p^2}{2m}+U(\vec r)$, the evolution equations for a small time $\tau$ can be written down as
\begin{equation}
\arraycolsep=1mm
    \left\{\begin{array}{ccccc}
      \delta^{(E)} x_i&=&\tau\cfrac{\partial E}{\partial p_i}&=&\tau \cfrac{p_i}{m}    \\\\
      \delta^{(E)} p_i&=&-\tau\cfrac{\partial E}{\partial x_i}&=&-\tau \nabla_i U(\vec r)   
    \end{array}.\right.   
\label{clevoE}\end{equation}
In sec.\ref{sec:Lop}, we introduced the angular momentum operator $\hat L$ and identified it with the evolution operator of the system under rotation over a small angle $\phi$. 
The corresponding evolution equations for a classical system with the angular momentum
\footnote{To write down a vector product (which enters the expression for the angular momentum) in tensor notations, one needs the 3d fully antisymmetric tensor, whose components in any right-handed Cartesian system equal $\varepsilon_{kij}=-\varepsilon_{kji}$, $\varepsilon_{kij}=\varepsilon_{ijk}=\varepsilon_{jki}$, $\varepsilon_{123}=1$.}  
$L_k=\varepsilon_{kij}x_ip_j$ can be written down as
\begin{equation}
\arraycolsep=1mm
    \left\{\begin{array}{ccccc}
      \delta^{(L)}_k x_i&=&\varphi\cfrac{\partial L_k}{\partial p_i}&=&-\varphi\,\varepsilon_{kij}x_j    \\\\
      \delta^{(L)}_k p_i&=&-\varphi\cfrac{\partial L_k}{\partial x_i}&=&-\varphi\,\varepsilon_{kij}p_j
    \end{array}.\right.   
\label{clevoL}\end{equation}
Note that the both relations (\ref{clevoE}, \ref{clevoL}) preserve relation between the coordinate and momentum due to
$\delta p_i=m\cfrac{d\delta x_i}{dt}$. 

What does notation (\ref{clevoE}, \ref{clevoL}) gives us? First, it follows that if evolution in time (\ref{clevoE}) preserves the angular momentum $L$, then the evolution corresponding the the operator $\hat L$ preserves the energy $E$, and vice versa,
\begin{equation}
    \frac{\delta^{(L)}_kE}{\varphi}=
    \cfrac{\partial E}{\partial x_i}\frac{\delta^{(L)}_kx_i}{\varphi}+
    \cfrac{\partial E}{\partial p_i}\frac{\delta^{(L)}_kp_i}{\varphi}=\cfrac{\partial E}{\partial x_i}\cfrac{\partial L_k}{\partial p_i}-\cfrac{\partial E}{\partial p_i}\cfrac{\partial L_k}{\partial x_i}=
    \cfrac{\partial L_k}{\partial x_i}\frac{\delta^{(E)}x_i}{\tau}+\cfrac{\partial L_k}{\partial p_i}\frac{\delta^{(E)}p_i}{\tau}=\frac{\delta^{(E)}L_k}{\tau}.
\end{equation}
The identities
$\delta^{(E)}E=0$ and $\delta_k^{(L)}L_k=0$ (no sum over $k$) are derived similarly. It is also easy to show that $\delta_k^{(L)}L^2=0$. On the other hand, the $3d$ motion, where both the energy and the angular momentum are preserved, can be fully determined by system of the three equations, $E(\vec r,\vec p)=C_1$, $L_k(\vec r,\vec p)=C_2$, $L^2(\vec r,\vec p)=C_3$ (where the right parts contain the constants of the corresponding dimension, and $k$ is any momentum component). Hence, if $\delta^{(L)}_kE=\delta^{(E)}L_k=0$ (for all components $k$), then the transformations (\ref{clevoE}) and (\ref{clevoL}) preserve the full set of the equations and in this sense are symmetries of the system. Hence,
\begin{center}
    \begin{tabular}{cp{15cm}}
        $\bullet$& If the energy and the angular momentum are conserved in a classical system, than each of these quantities generates a symmetry of the system given by (\ref{clevoE}, \ref{clevoL}).
    \end{tabular}
\end{center}

\paragraph{Symmetry for the LRL vector.}
Now we have one more way to establish a relation between the conserving quantities and symmetries of the system.
Lets apply it to the new conserving quantity, which is the LRL vector, which unlike the energy and angular momentum is not \textit{a priory}
related to any symmetries of the system. Consider the transformation
\begin{equation}
\arraycolsep=1mm
    \left\{\begin{array}{ccccc}
      \delta^{(A)}_k x_i&=&\alpha\cfrac{\partial A_k}{\partial p_i}&=& 2p_i x_j - x_i p_j-\delta_{ij}(\vec r, \vec p)    \\\\
      \delta^{(A)}_k p_i&=&-\alpha\cfrac{\partial A_k}{\partial x_i}&=&  \dfrac{p_i p_j}{m} - k \dfrac{x_i x_j}{r^3}-\dfrac{p^2}{m}\delta_{ij}+\dfrac{k}{r}\delta_{ij}  
    \end{array}.\right. \label{clevoA}
\end{equation}
One can check intermediately that $m\cfrac{\delta^{(A)}_k x_i}{dt}=\delta p_i$, i.e., the relation between the coordinate and the momentum is preserved. Then reasoning as above, we obtain
$\delta^{(A)}_kA_k=0$ (identically); $\delta^{(A)}_kE_k\sim\delta^{(E)}_kA_k=0$ (because the $\vec A$ is conserved, i.e., does not change under the evolution in time); 
$\delta^{(A)}_kL_k=0$ (checked by a straightforward computation). Then the transformation generated by $A_k$ according to (\ref{clevoA}) conserves the full set of the equations of motions in the form $E(\vec r,\vec p)=C_1$, $A_k(\vec r,\vec p)=C_1$,  $L_k(\vec r,\vec p)=C_3$ ($C$ are dimension constants, $k$ is any component) and hence are symmetries of the system. Thus
\begin{center}
    \begin{tabular}{cp{15cm}}
        $\bullet$& If the energy, the angular momentum component $L_k$ and the LRL vector component $A_k$ are preserved in a classical system, then each of the quantities generates symmetry of the system (\ref{clevoE}, \ref{clevoL},\ref{clevoA}).
    \end{tabular}
\end{center}
Form (\ref{clevoE},\ref{clevoL},\ref{clevoA}) of the time shift and other transformations of a classical system is called their \textit{hamiltonian} form. 

\paragraph{Classical analogue of the simultaneous(im)measurability.}
Above we considered a curious property of writing down the equation of motion (the transformations in time) and the other transformations of the system in hamiltonian form (\ref{clevoE},\ref{clevoL},\ref{clevoA}), which generally can be written down as
\begin{center}
\begin{tabular}{cp{15cm}}
$\bullet$& Change of the quantity $F$ w.r.t. the small transformation generated by the quantity $G$, as well as the change of teh quantity
$G$ under the small transformation generated by the quantity $F$ equals
$\delta^{(F)}G=\delta^{(G)}F\equiv\alpha\{F,G\}$, where $\alpha$ is a small transformation parameter.
\end{tabular}
\end{center}
Bilinear antisymmetric function $\{F,G\}$ is called a \textit{Poisson bracket} of the quantities $F$ and $G$. 

A sense of the Poisson brackets is much more transparent, if one considers their quantum analogue, which we are in fact already familiar from sec.\ref{sec:Lop}. We introduced there the components of the angular momentum operator $\hat L_x$, $\hat L_y$, $\hat L_z$ and identified them with the operators of the infinitesimal rotations around the corresponding axis. 
The simultaneous immeasuarability of different components was then related to nonvanishing of their commutators, $[L_x,\hat L_y]\ne 0$, $[L_x,\hat L_z]\ne 0$, $[L_y,\hat L_z]\ne 0$. On the contrary, the simultaneous measuarability of each component with the square momentum operator was related to vanishing of the commutators
$[L^2,\hat L_x]=[L^2,\hat L_y]=[L^2,\hat L_z]= 0$. (Recall that the commutator of the operators $\hat F$ and $\hat G$ is by definition the bilinear combination  $[\hat F,\hat G]\equiv \hat F\hat G-\hat G\hat F$). It turns out that a commutator in quantum theory has the same sense as the Poisson brackets in classical theory,
\begin{equation}
    \begin{array}{lc}
    \multicolumn{2}{p{8.5cm}}{Changing of F and G under the transformation, generated by G or F ($\alpha$ is a small parameter)}\\\\
    \text{Classically}&\delta^{(F)}G=\delta^{(G)}F=\alpha\{F,G\}\\
    \text{}&\delta^{(\hat F)}\hat G=\delta^{(\hat G)}\hat F=\alpha [\hat F,\hat G]
    \end{array}
\end{equation}
We have already touched on this in sec.\ref{sec:Lop}, and we will return to it in sec.\ref{sec:qLRL}.

The analogy between the Poisson brackets in the commutator goes futher. In quantum theory, vanishing of the pairwise commutators for a set of operator observables means that \textit{all} these operators can be used \textit{at the same time} to construct a set of the quantum numbers that describes a given state (see sec.\ref{sec:multrad}, \ref{sec:qLRL}). In classical theory, vanishing of pairwise Poisson brackets of a set of quantities means that \textit{all} these quantities can be used \textit{at the same time} to simplify the equations of motion by substituting each such quantity for a pair coordinate--impulse \cite{Arn}. If there are as many pairwise commutating operator observables as coordinates, then the corresponding set of quantum numbers fully describes the state and is called a \textit{full set} \cite{LL3}. If there are as many classical quantities with the pairwise vanishing Poisson brackets as coordinates, these quantities are enough to substitute with them all coordinates and momenta. The system then gets especially simple description in terms of such quantities and is called \textit{completely integrable system} \cite{Arn}.

\subsection{Quantum analogue of the RLR vector\label{sec:qLRL}}
\paragraph{Conservation of the operator observables.}
As we discussed in sec.\ref{sec:RBLight}, in quantum mechanics a state of the system is described by a vector of the state space (denote if $\Psi$ here) and by the evolution operator. Moreover, one can introduce the infinitesimal evolution operator,
\begin{equation}
\delta^{(H)}\Psi=\tau\hat H\Psi,\qquad \tau\to0. \label{qevoH}
\end{equation}
In sec.\ref{sec:Lop}, we used the example if the infinitesimal rotation operator to discuss, the evolution of the system w.r.t. space transformations can be described in the same way,
\begin{equation}
    \delta^{(L)}_k \Psi=\varphi\hat L\Psi.\label{qevoL}
\end{equation}
Let's find out, when transformation (\ref{qevoL}) preserves evolution equation (\ref{qevoH}). Rewrite (\ref{qevoH}) before the rotation as
\begin{equation}
\tilde\Psi=\Psi+\tau\hat H\Psi\label{qevoH1}
\end{equation}
Under a rotation,
\begin{equation}
    \Psi\to \Psi+\varphi\hat L \Psi,\qquad \tilde\Psi\to\tilde\Psi+\varphi\hat L\tilde\Psi,\label{qevoL1}
\end{equation}
and it must be that
\begin{equation}
\Psi+\tau\hat H\Psi+\varphi\hat L\big(\Psi+\tau\hat H\Psi\big)
\stackrel{(\ref{qevoH1})}{=}\tilde{\Psi}+\varphi\hat L\tilde{\Psi}\stackrel{(\ref{qevoL1})}{=}
\Psi+\varphi\hat L\Psi+\tau\hat H\big(\Psi+\varphi\hat L\Psi\big),
\end{equation}
whence
\begin{equation}
    \big(\hat L\hat H-\hat H\hat L\big)\Psi\equiv[\hat L,\hat H]\Psi=0.
\end{equation}
If the above is true for any vector $\Psi$ of the space state, then 
$[\hat L,\hat H]=0$. This is the condition that (\ref{qevoL}) preserves the evolution equation (\ref{qevoH}). One can reverse the logic and consider (\ref{qevoL}) as evolution equation of the system under a rotation and (\ref{qevoH}) as transformation of the system under the time shift. 
Then a similar argument gives $[\hat L,\hat H]=0$, already as the condition that transformation (\ref{qevoH}) preserves equation (\ref{qevoL}). One can also say that the \textit{operator} $\hat L$ \textit{is conserved} in time and the the operator $\hat H$ is conserved under rotations, meaning the conservation of equation  (\ref{qevoL}) and (\ref{qevoH}) under the corresponding transformations.
Hence,
\begin{center}
\begin{tabular}{cp{15cm}}
$\bullet$&The operator of the infinitesimal evolution $H$ is preserved under rotations if and only if the momentum opeator, which has the form of the infinitesimal rotation operator, is preserved in time, if and only if $[\hat L,\hat H]=0$.
\end{tabular}
\end{center}

\paragraph{Evolution operator of the electron in atom.}
In sec.\ref{sec:trans}, we got familiar with notation of the evolution operator via the eigenvectors and eigenvalues,
\begin{equation}
    \hat H\Psi^{(\lambda)}=\lambda\Psi^{(\lambda)},\qquad \delta^{(H)}\Psi^{(\lambda)}=\tau\lambda\Psi^{(\lambda)},\label{evoHei}
\end{equation}
which (in case of time-independent $\hat H$) allows one to easily write the solution for $\Psi(t)$
via expansion over $\Psi^{(\lambda)}(t)$. The above condition $[\hat H,\hat L]$ for (\ref{evoHei}) means the the function $\tilde{\Psi}=\Psi^{(\lambda)}+\varphi\hat L_k\Psi^{(\lambda)}$ transformed by the operator $\hat L$  will be solution as well and with the same $\lambda$. Moreover, a well known theorem of linear algebra \cite{Gel} allows one to find a basis of the eigenvectors of the  $\hat L_k$ on the subspace spanned over the images of the vector $\Psi^{(\lambda)}$ under the multiple action of the operator $\hat L_k$. I.e., one can describe the eigenvectors in more details, enumerating them by the two eigenvalues (``quantum numbers''),
\begin{equation}
    \hat H\Psi^{(\lambda,\mu)}=\lambda\Psi^{(\lambda,\mu)},
    \qquad \hat L_k\Psi^{(\lambda,\mu)}=\mu\Psi^{(\lambda,\mu)}.
\end{equation}
If there is one more operator, say $L^2\equiv \hat L_x^2+\hat L_y^2+\hat L_z^2$, such that $[\hat L^2,\hat H]=[\hat L^2,\hat L_k]=0$, then one can write in the above equations $\Psi^{(\lambda,\mu,\nu)}$ completing them with one more equation $\hat L^2\Psi^{(\lambda,\mu,\nu)}=\nu\Psi^{(\lambda,\mu,\nu)}$.
This is what we in fact did in sec.\ref{sec:multrad}, where we described an electron state by the quantum numbers
$(n,l,m)$, although we did not write the operator $\hat H$ explicitly. We see that
\begin{center}
\begin{tabular}{cp{15cm}}
$\bullet$&The triple of the operators $(\hat H,\hat L_k,\hat L^2)$ that is necessary and sufficient for the complete description of the states of quantum system via the quantum numbers corresponds the the triple of the conserving quantities $(E,L_k,L^2)$ that is necessary and sufficient to write down the complete set of equations of motion for the classical system (see sec.\ref{sec:clRLR}).
\end{tabular}
\end{center}
For the classical system, we found the LRL conservation law, which allowed as to find an alternative 
 triple of the equations of motion that corresponds the the conserving quantities $(E,L_k,A_k)$.
Hence there is a question, whether there is an operator analogue of the LRL vector for a quantum system, i.e., the operator $\hat A$, such that $[\hat A,\hat H]=[\hat A,\hat L_k]=0$? As we see in sec.\ref{sec:SO4}, the answer is positive. Moreover, it turns out that $[\hat A,\hat L^2]\ne 0$. And now let's see, what will give us such an operator if any.

\paragraph{Simultaneous measuarability and accidental degeneration.}
Hence, the electron state in the atom can be described by one of the two triples of quantum numbers
\begin{equation}
    \hat H\Psi^{(\lambda,\mu,\nu)}=\lambda\Psi^{(\lambda,\mu,\nu)},
    \qquad \hat L_k\Psi^{(\lambda,\mu,\nu)}=\mu\Psi^{(\lambda,\mu,\nu)},\qquad \hat L^2\Psi^{(\lambda,\mu,\nu)}=\nu\Psi^{(\lambda,\mu,\nu)},
\end{equation}
\textbf{or}
\begin{equation}
    \hat H\Psi^{(\lambda,\mu,\vartheta)}=\lambda\Psi^{(\lambda,\mu,\vartheta)},
    \qquad \hat L_k\Psi^{(\lambda,\mu,\vartheta)}=\mu\Psi^{(\lambda,\mu,\vartheta)},\qquad \hat A_k\Psi^{(\lambda,\mu,\vartheta)}=\vartheta\Psi^{(\lambda,\mu,\vartheta)}.
\end{equation}
This is because $[\hat H,\hat L_k]=[\hat H,\hat L^2]=[\hat H,\hat A_k]=[\hat L^2,\hat L_k]=[\hat A_k,\hat L_k]=0$, but $[\hat L^2,\hat A_k]\ne 0$. Due to the same relations, a state transformed by the operator $\hat A_k$,
 $\tilde{\Psi}^{(\hat A_k)}\equiv\Psi{(\lambda,\mu,\nu)}+\alpha\hat A_k\Psi{(\lambda,\mu,\nu)}$, satisfies
\begin{equation}
    \hat H\tilde{\Psi}^{(\hat A_k)}=\lambda\tilde{\Psi}^{(\hat A_k)},\qquad \hat L_k\tilde{\Psi}^{(\hat A_k)}=\mu\tilde{\Psi}^{(\hat A_k)},
\end{equation}
but generally,
\begin{equation}
    \hat L^2\tilde{\Psi}^{(\hat A_k)}\ne\nu\tilde{\Psi}^{(\hat A_k)}.
\end{equation}
In other words, because $\Psi^{\lambda,\mu,\nu}$ form a basis in the state space,
    \begin{equation}    
    \tilde{\Psi}^{(\hat A_k)}=C_{\nu}\Psi^{\lambda,\mu,\nu}+\sum_{\nu'\ne\nu}C_{\nu'}\Psi^{\lambda,\mu,\nu'},
    \end{equation} 
where, generally, at least one $C_{\nu'}$ is other than $0$. 
Hence,
\begin{center}
\begin{tabular}{cp{15cm}}
     $\bullet$ & There is more than one state (and then more than one-dimensional state space) with one and the same eigenvalues of $\hat H$ and $\hat L_k$, but with different eigenvalues of $\hat L^2$ (the latter ensures linear independence of the states).
\end{tabular}
\end{center}
This fact is called \textit{accidental degeneracy} of the Coulomb spectrum. It happens thanks to the LRL operator, which ``conserves'' under the action of the infinitesimal evolution operator $\hat H$ but does not allow one to complete the set of quantum numbers that describes the system without the LRL vector (but allows one to substitute this set for the alternative one).

\subsection{Quantum symmetries of the Coulomb problem\label{sec:SO4}}
\paragraph{Explicit form of the evolution and LRL operators.}
In sec.\ref{sec:multrad}, we mentioned Scr\"odinger equation for a stationary state of an atom, which is described by the function $\psi(r,\theta,\varphi)$. Now it is time to give its explicit expression for the simplest (hydrogen) atom \cite{LL3}: 
\begin{equation} \label{Sch}
-\frac{h^2}{2\mu}\Delta\psi-\frac{k}{r}\phi=E\psi       
\end{equation}
Without the term $-\frac{k}{r}$, this equation coincides with the Helmholz equation for the classical monochromatic radiation (see. sec.\ref{sec:multrad}). The term $-\frac{k}{r}$ accounts for the potential energy of the electron in the Coulomb nucleus field. The $E$ is here the full energy of the electron.
Comparing  (\ref{Sch}) with eigenvalue form of the evolution equation (\ref{evoHei}), we obtain that the infinitesimal evolution operator equals 
\begin{equation}
    \hat H=-\frac{h^2}{2\mu}\Delta-\frac{k}{r}.\label{Hsch}
\end{equation}
Now we try to guess the expression for the LRL vector. In sec.\ref{sec:QclL}, we discussed the correspondence of the classical angular momentum and quantum operator of the angular momentum\footnote{Unlike sec.\ref{sec:mulSch}, here we use the common normalisation of the angular momentum operator, see the footnote at the beginning of sec.\ref{sec:Lop}.}. Moreover, comparing the corresponding equations, one can suggest an operator corresponding to the classical momentum,
\begin{equation}
L=\varepsilon_{ijk}x_ip_i\ \leftrightarrow\ \hat L=\hbar\varepsilon_{ijk}x_i\partial_i,\qquad
p_i\ \leftrightarrow\ \hat p_i=\hbar\partial_i.\label{QclLP}
\end{equation}
Substituting these correspondences in the expression for the LRL vector in the simplest way, we get
\begin{equation} \label{ope}
\hat A_i    = \frac
{\varepsilon_{ijk} \hat p_i \hat L_k}{m} - k\frac{x_i}{r}.
\end{equation} 
In standard quantum mechanics formalism the operator observable must be Hermitian \cite{LL3}, $\hat A^{dagger}=\hat A$. In particular, the operator then generates a basis of the eigenvectors\footnote{Here $\dagger$ is the Hermitian conjugation, which is the complex conjugation + the transposition \cite{Gel}).} \cite{Gel}.
To achieve that, substitute the above naive expression with with the its simplest generalisation,
$\frac{1}{2}(\hat p_i \hat L_k + \hat L_k \hat p_i) $. Then instead of (\ref{ope}),  we get
   \begin{equation} \label{ope1}
\hat A_i    = \varepsilon_{ijk} \frac{1}{2m}(\hat p_i \hat L_k + \hat L_k \hat p_i)  - k\frac{x_i}{r}.
\end{equation} 
This operator is already hermitian. One can straightforwardly check that  $[\hat A_i,\hat{H}] = 0$ and $[\hat A_i,L_i]=0$ (see the Appendix). Hence, we have found the needed operator (see sec.\ref{sec:qLRL} at the end). Thus
\begin{center}
    \begin{tabular}{cp{15cm}}
        $\bullet$&One can use the triple of the operators $(\hat H,\hat L_k,\hat A_k)$ given by (\ref{Hsch},\ref{QclLP},\ref{ope1}) to describe the quantum state of the electron in the hydrogen atom by an alternative triple of quantum numbers, which are eigenvalues of the operators made dimensionless by the system constants.
    \end{tabular}
\end{center}

\paragraph{$4d$ rotations.}
Note that $[\hat A_i,\hat{H}] = 0$ means that also $[[\hat A_i,\hat A_k],\hat{H}]=0$, and, due to the relations $ [\hat L_k,\hat{H}] = 0$, also $ [[\hat A_i,\hat L_k]\hat{H}] = 0$, and one can proceed so on. Thus we obtained the entire system of the operators commutating with $\hat H$ and hence representing the symmetries of our problem. How large is this system, and how is it constructed? To answer the question, calculate the commutators (see the Appendix), 
\begin{equation} \label{commAL}
[\hat L_i,\hat A_j]=i\hbar\varepsilon_{ijk}\hat A_k,\qquad
[\hat L_i,\hat L_j]= i\hbar\varepsilon_{ijk}\hat L_k,\qquad
[\hat A_i,\hat A_j] = -2i\hbar\hat{H} \varepsilon_{ijk} \hat L_k.  
\end{equation}
Consider the subspace spanned over the eigenvectors of the operator $\hat{H}$ with a given value of the energy $E$ in (\ref{Sch}). On this subspace, one can substitute $\hat{H}$ with $E$ in relations (\ref{commAL}) and introduce the operators 
$\hat{u_i} = \frac{\hat A_i}{\hbar\sqrt{-2E}}$, $\hat l_i=\frac{\hat L^i}{\hbar}$. Then relations (\ref{commAL}) are rewritten as  
\begin{equation} \label{commlu}
[\hat L_i,\hat{u_j}]=i\varepsilon_{ijk}\hat{u_k},\qquad
[\hat L_i,\hat L_j]= i\hbar\varepsilon_{ijk}\hat L_k,\qquad
[\hat{u_i},\hat{u_j}] = i\varepsilon_{ijk} \hat L_k.  
\end{equation}

The same commutation relations are satisfied by the matrices of the infinitesimal rotations in the $4d$ space,
\begin{equation}
\begin{array}{c}
\Omega_{zt} = \left(\begin{array}{cccc}
0 & 0 & 0 & 0\\
0 & 0 & 0 & 0\\
0 & 0 & 0 & -1\\
0 & 0 & 1 & 0
\end{array}\right)	,\qquad
\Omega_{yt} = \left(\begin{array}{cccc}
0 & 0 & 0 & 0\\
0 & 0 & 0 & 1\\
0 & 0 & 0 & 0\\
0 & -1 & 0 & 0
\end{array}\right)  ,\qquad
\Omega_{xt} = \left(\begin{array}{cccc}
0 & 0 & 0 & 1\\
0 & 0 & 0 & 0\\
0 & 0 & 0 & 0\\
-1 & 0 & 0 & 0
\end{array}\right),\\\\
\Omega_{xy} = \left(\begin{array}{cccc}
0 & 1 & 0 & 0\\
-1 & 0 & 0 & 0\\
0 & 0 & 0 & 0\\
0 & 0 & 0 & 0
\end{array}\right)	,\qquad
\Omega_{xz} = \left(\begin{array}{cccc}
0 & 0 & 1 & 0\\
0 & 0 & 0 & 0\\
-1 & 0 & 0 & 0\\
0 & 0 & 0 & 0
\end{array}\right)	 ,\qquad
\Omega_{yz} = \left(\begin{array}{cccc}
0 & 0 & 0 & 0\\
0 & 0 & -1 & 0\\
0 & 1 & 0 & 0\\
0 & 0 & 0 & 0
\end{array}\right), 
\end{array}\label{4drot}
\end{equation}
if one puts in the correspondence for $(i,j,k)\in\{x,y,z\}$: $\Omega_{ij}\leftrightarrow\varepsilon_{ijk}\hat L_k$ and $\Omega_{it}\leftrightarrow \hat A_i$.
In sec.\ref{sec:symmult}, we discussed representations of the 3d rotation group, i.e., maps of 3d rotations to linear operators than preserves the composition. Now we encounter with one more way to construct a representation of the rotation group. Namely, one can construct a map of the infinitesimal rotations into the linear operators that preserves the commutation relations \cite{Vil}. Both (\ref{commlu}) and (\ref{4drot}) are examples of such maps for the $4d$ rotation group. To construct the representation of the rotation group (the group elements themselves), one can compute the operator (matrix) exponentials (see. sec.\ref{sec:evoexp}) from $\hat \Omega_{ij}$ and all their commutators (one representation), or do in the same with $\hat l_i$, $\hat u_i$ (another representation).

\subsection{Perihelion precession and the Lamb shift}
Let us emphasise that the above classical conservation law of the LRL vector and the ``conserving'' quantum LRL operator, as well as the classical and quantum symmetries behind them -- are the unique properties of the $\frac{k}{r}$ potential. If the potential is at least a little different from $\frac{k}{r}$, in particular if it has arbitrary small correction of the form  $\frac{a}{r^3}$, everything breaks. 

In classical problem, a correction of the form $\frac{a}{r^3}$ to the potential energy of the planet leads to that the orbit is no longer closed (see sec.\ref{sec:clRLR}). The planets moves then in a spiral similar to a slowly rotating ellipse \cite{Arn}. This phenomenon is known as the perihelion precession. In case of real planet motion, such correction is predicted by the general relativity (GR). Hence, comparing the observed perihelion precession of the Mercury with the theoretical prediction of the GR became one of the experimental checks of the GR \cite{MerPer}. The perihelion precession as a phenomenon still plays an important role in this science \cite{GenRel}.

In quantum problem, the $\frac{a}{r^3}$ correction to the potential energy of the electron in a hydrogen atom (to the term $\frac{k}{r}$ in the Schr\"odinger equation) leads to removal of the accidental degeneracy (see sec.\ref{sec:clRLR}). I.e. the states that had the same energy (the eigenvalue of the operator $\hat H$) now have different energies. This is observed as a split of one spectral line to several ones. Such split in the hydrogen spectrum is predicted by the quantum electrodynamics (QED) \cite{FeyInt}. The precision measurement of the magnitude of this split confirm with a high precision the prediction of QED\cite{Lamb1,Lamb2,Lamb3}.

\section*{Acknowlegements}
A.Anokhina is grateful to E.Vyrodov for reading the draft and making useful comments.

\

The work of A.Anokhina was supported by the Russian Science Foundation (grant № 23-41-00049).

\printbibliography


\appendix

\section{Commutator $[\hat A_i,\hat{H}]$}

$$\hat{A_x} = \frac {\hat{p_y} \hat{L_z} - \hat{p_z} \hat{L_y} - \hat{L_y} \hat{p_z} - \hat{L_z} \hat{p_y}}{2m} -\frac{kx}{r} ,\qquad \hat{H} = -\frac{h^2}{2m} \Delta - \frac{k}{r}$$
  
\begin{equation} \label{CommAxH}
[\hat{A_x},\hat{H}] = \underbrace{\frac {[\hat{p_y},\hat{H}] \hat{L_z} - [\hat{p_z},\hat{H}] \hat{L_y} - \hat{L_y} [\hat{p_z},\hat{H}]- \hat{L_z} [\hat{p_y},\hat{H}]}{2m}}_{\equiv F} -\underbrace{k[\frac{x}{r},\hat{H]}}_{\equiv G}.
\end{equation}

\begin{equation} \label{F1}
[\hat p_i,\hat{H}] = -ihk\frac{{x_i}}{r^3}\ \Rightarrow\
F = \textstyle{\frac{-e H^2}{2m} \left(\frac{y}{r^3} (x \frac{\partial}{\partial y} - y \frac{\partial}{\partial x} ) - \frac{z}{r^3} (z \frac{\partial}{\partial x} - x \frac{\partial}{\partial z} ) -  (z \frac{\partial}{\partial x} - x \frac{\partial}{\partial z} ) \frac{z}{r^3} -  (x \frac{\partial}{\partial y} - y \frac{\partial}{\partial x} ) \frac{y}{r^3}\right)}
\end{equation}

$$
    \textstyle{(z \frac{\partial}{\partial x} - x \frac{\partial}{\partial z} ) \frac{z}{r^3} = \frac{z^2}{r^3} \frac{\partial}{\partial x} - \frac{x}{r^3} - \frac{xz}{r^3} \frac{\partial}{\partial z},\qquad
    (x \frac{\partial}{\partial y} - y \frac{\partial}{\partial x} ) \frac{y}{r^3} = -(\frac{y^2}{r^3} \frac{\partial}{\partial x} - \frac{x}{r^3} - \frac{xy}{r^3} \frac{\partial}{\partial y })},\qquad \Rightarrow
$$

\begin{equation} \label{F2}
    \Rightarrow\qquad F = -\frac{kh^2}{2m} \left(\frac{xy}{r^3}\frac{\partial}{\partial y} + \frac{xz}{r^3}\frac{\partial}{\partial z} + \frac{x}{r^3} - \frac{y^2 + z^2}{r^3}\frac{\partial}{\partial x}\right).
\end{equation}

\begin{equation} \label{G}
    G = \frac{-kh^2}{2m} [\frac{x}{r},\Delta ]
    = \frac{-kh^2}{m} \left(-\frac{x}{r^3} + \frac{1}{r} \frac{\partial}{\partial x} - \frac{x^2}{r^3} \frac{\partial}{\partial x}-\frac{xy}{r^3} \frac{\partial}{\partial y}-\frac{xz}{r^3} \frac{\partial}{\partial z}\right).
\end{equation}

$$\Rightarrow\qquad \boxed{[\hat A_i,\hat{H}]=F-G=0}.$$

\section{Commutator $\left[\hat{L_\alpha},\hat A_i\right]$}


Write down the LRL operator explicitly.
$$\hat A_i = \frac{1}{2m} \left(\varepsilon_{ijk}\hat p_j\hat L_k - \varepsilon_{ikj}\hat L_k\hat p_j\right) - k \frac{x_i}{r} =$$ $$= -\frac{\hbar^2}{2m} \varepsilon_{ijk}\varepsilon_{kmn}\left( \frac{\partial}{\partial x_j}x_m\frac{\partial}{\partial x_n} + x_m\frac{\partial}{\partial x_n}\frac{\partial}{\partial x_j}\right)  - k \frac{x_i}{r} =$$
$$ = -\frac{\hbar^2}{2m} \left(\delta_{im}\delta_{jn} - \delta_{in}\delta_{jm}\right)\left(2x_m\frac{\partial^2}{\partial x_j\partial x_n}+\delta_{jm}\frac{\partial}{\partial x_n}\right)   - k \frac{x_i}{r} = $$$$ = \frac{\hbar^2}{m} \left(-x_i\frac{\partial^2}{\partial x^2_n} + 2x_m\frac{\partial^2}{\partial x_n\partial x_i} + \frac{\partial}{\partial x_n}\right)  - k \frac {x_i}{r}  $$
Substitute the expression to the commutation relations.
$$\left[\hat{L_\alpha},\hat A_i\right] = \left[\varepsilon_{\alpha j k} x_{j} p_{k}, \frac{\hbar^2}{m} \left( \frac{\partial}{\partial x_i} + x_n \frac{\partial^2}{\partial x_n \partial x_i} - x_i \frac{\partial^2}{\partial x^2_n}\right) - k \frac {x_i}{r}\right] =  $$
 $$ = i\frac{\hbar^3}{m} \varepsilon_{\alpha j k} \left( -\left[x_j \frac{\partial}{\partial x_k},\frac{\partial}{\partial x_i }\right] - \left[x_j \frac{\partial}{\partial x_k}, x_n  \frac{\partial^2}{\partial x_n \partial x_i}\right] + \left[x_j \frac{\partial}{\partial x_k} , \frac{\partial^2}{\partial x^2_n}\right]\right) + i \hbar k \varepsilon_{\alpha j k} \left[x_j \frac{\partial}{\partial x_k}, \frac{x_i}{r}\right]  =  $$

$$ = i \frac{\hbar^3}{m} \varepsilon_{\alpha j k} \left( \delta_{ij}  \frac{\partial}{\partial x_k} - x_j \frac{\partial^2}{\partial x_k\partial x_i} - x_j x_n \frac{\partial^3}{\partial x_n\partial x_i\partial x_i} +x_n \frac{\partial}{\partial x_n} \left( \delta _{ij} \frac{\partial}{\partial x_k} + x_j \frac{\partial^2}{\partial x_i\partial x_k} \right)    + x_j \delta_{ik} \frac{\partial^2}{\partial x^2_n} +  \right.$$ $$\left.  + x_i x_j \frac{\partial^3}{\partial x^2_n\partial x_k}  - x_i \frac{\partial}{\partial x_n} \left(\delta_{jn} \frac{\partial}{\partial x_k} + x_j \frac{\partial^2}{\partial x_k\partial x_n} \right) \right)  + i \hbar k \varepsilon_{\alpha j k} \left(x_j \delta_{ik} \frac{1}{r} - \frac {x_i x_j x_k}{r^3} \right)    =   $$
$$ = i\frac{\hbar^3}{m} \varepsilon_{\alpha jk} \left(  \delta_{ij} \frac{\partial}{\partial x_k} - x_j \frac{\partial^2}{\partial x_k\partial x_i} -  x_j x_n \frac{\partial^3}{\partial x_k\partial x_n\partial x_i} + \delta_{ij} x_n \frac{\partial^2}{\partial x_n\partial x_k}    + x_j \frac{\partial^2}{\partial x_i\partial x_k} + x_n x_j \frac{\partial^3}{\partial x_n\partial x_i \partial x_k} +\right.$$ $$ \left.+ x_j \delta_{ik} \frac{\partial^2}{\partial x^2_n} - x_i \frac{\partial^2}{\partial x_j \partial x_k }   - x_i \frac{\partial^2}{\partial x_k \partial x_j } - x_i x_j \frac{\partial^3}{\partial x^2_n \partial x_k } + x_i x_j \frac{\partial^3}{\partial x^2_n \partial x_k }\right) + i \hbar k \varepsilon_{\alpha j k} \left(x_j \delta_{ik} \frac{1}{r} - \frac {x_i x_j x_k}{r^3} \right) =   $$

$$ = -i \frac{\hbar^3}{m} \varepsilon_{\alpha j k } \left( \delta_{ij} \frac{\partial}{\partial x_k} + \delta_{ij} x_n \frac{\partial^2}{\partial x_n\partial x_k} - x_k \delta_{ij} \frac{\partial^2}{\partial x^2_n} - 2 x_i \frac{\partial^2}{\partial x_j\partial x_k} \right) + i \hbar k \varepsilon_{\alpha j k} \left(x_j \delta_{ik} \frac{1}{r} - \frac {x_i x_j x_k}{r^3} \right) =       $$
Because $\varepsilon_{\alpha j k }$ is the fully antisymmetric tensor, its contraction with the symmetric tensors is $0$.  Hence we obtain

   $$  =  i \hbar \varepsilon_{\alpha j k } \delta_{ij}\left[ \frac{\hbar^2}{m} \left(  \frac{\partial}{\partial x_k}  + x_n \frac{\partial^2}{\partial x_n\partial x_k} - x_k \frac{\partial^2}{\partial x^2_n}\right) -k \frac{x_k}{ r } \right]  = ih \varepsilon_{\alpha i k } \hat A_k $$

\section{Commutator $\left[\hat A_i,\hat A_j\right]$}

$$ \left[\hat A_i,\hat A_j\right] = \left[\left( k \frac{x_i}{ r } + \frac{\hbar^2}{m} \left(  -\frac{\partial}{\partial x_i}  - x_m \frac{\partial^2}{\partial x_m\partial x_i} + x_i \frac{\partial^2}{\partial x^2_m}\right)  \right), \left( k \frac{x_j}{ r } +\frac{\hbar^2}{m} \left(  -\frac{\partial}{\partial x_j}  - x_k \frac{\partial^2}{\partial x_k\partial x_j} + x_j \frac{\partial^2}{\partial x^2_k}\right)  \right)\right]  =    $$

Decompose the expression

$$  \frac{x_i}{r}     x_j \frac{\partial^2}{\partial x^2_k}     x_k \frac{\partial^2}{\partial x_k \partial x_j}         \frac{\partial}{\partial x_j}        $$

$$  k^2 \left[\frac{x_i}{r}, \frac{x_j}{r}  \right] + \frac{k}{m} \hbar^2 \left[\frac{x_i}{r}, x_j \frac{\partial^2}{\partial x^2_k} \right] - \frac{k}{m} \hbar^2 \left[\frac{x_i}{r}, x_k \frac{\partial^2}{\partial x_k \partial x_j}\right] -\frac{k}{m} \hbar^2 \left[\frac{x_i}{r}, \frac{\partial}{\partial x_j}\right] +\frac{k}{m} \hbar^2 \left[x_i \frac{\partial^2}{\partial x^2_m},\frac{x_i}{r} \right] + $$

$$ + \frac{\hbar^4}{m^2} \left[x_i \frac{\partial^2}{\partial x^2_m},x_j \frac{\partial^2}{\partial x^2_k}  \right] - \frac{\hbar^4}{m^2} \left[x_i \frac{\partial^2}{\partial x^2_m}, x_k \frac{\partial^2}{\partial x_k \partial x_j} \right] - \frac{\hbar^4}{m^2} \left[x_i \frac{\partial^2}{\partial x^2_m},\frac{\partial}{\partial x_j} \right] -\frac{k}{m} \hbar^2 \left[x_m \frac{\partial^2}{\partial x_m \partial x_i},\frac{x_j}{r} \right] - $$ 

$$  - \frac{\hbar^4}{m^2} \left[x_m \frac{\partial^2}{\partial x_m \partial x_i}, 
 x_j \frac{\partial^2}{\partial x^2_k}\right] + \frac{\hbar^4}{m^2} \left[x_m \frac{\partial^2}{\partial x_m \partial x_i}, x_k \frac{\partial^2}{\partial x_k \partial x_j}\right] + \frac{\hbar^4}{m^2} \left[x_m \frac{\partial^2}{\partial x_m \partial x_i}, \frac{\partial}{\partial x_j} \right] - \frac{k}{m} \hbar^2 \left[\frac{\partial}{\partial x_i}, \frac{x_j}{r} \right] - $$
 
 $$ -\frac{\hbar^4}{m^2} \left[\frac{\partial}{\partial x_i}, x_j \frac{\partial^2}{\partial x^2_k} \right] +\frac{\hbar^4}{m^2} \left[\frac{\partial}{\partial x_i}, x_k \frac{\partial^2}{\partial x_k \partial x_j}\right] + \frac{\hbar^4}{m^2} \left[\frac{\partial}{\partial x_i},\frac{\partial}{\partial x_j}\right]  $$

One can easily see that the first and the last addends are zero.
Computing the commutator, we obtain the follows.

$$ \left[\frac{x_i}{r}, x_j \frac{\partial^2}{\partial x^2_k} \right] = 2\left( \frac{x_i x_j }{r^3} + \frac{x_i x_j x_k }{r^3}\frac{\partial}{\partial x_k} - \frac{x_j}{r}\frac{\partial}{\partial x_i}\right)  $$

 $$ \left[\frac{x_i}{r}, x_k \frac{\partial^2}{\partial x_k \partial x_j}\right] = \delta_{ij} \frac{1}{r} - \frac{x_i x_j }{r^3} +\left( \frac{x_i x_j x_k }{r^3} - \delta_{ij} \frac{x_k}{r}\right) \frac{\partial}{\partial x_k}   $$

 $$ \left[\frac{x_i}{r}, \frac{\partial}{\partial x_j}\right] = -\delta_{ij} \frac{1}{r} + \frac{x_i x_j }{r^3} $$
 
 $$\left[x_i \frac{\partial^2}{\partial x^2_m}, \frac{x_j}{r} \right] = -2\left( \frac{x_i x_j }{r^3} +  \frac{x_i x_j x_k }{r^3}\frac{\partial}{\partial x_k} -\frac{x_i}{r} \frac{\partial}{\partial x_j} \right)  $$
 
 $$\left[x_i \frac{\partial^2}{\partial x^2_m},x_j \frac{\partial^2}{\partial x^2_k}  \right] =      2\left(x_i \frac{\partial^3}{\partial^2 x_k \partial x_j} - x_j \frac{\partial^3}{\partial x^2_k \partial x_i}\right)
    $$
 
 $$\left[x_i \frac{\partial^2}{\partial x^2_m}, x_k \frac{\partial^2}{\partial x_k \partial x_j} \right]  = x_i \frac{\partial^3}{\partial^2 x_k \partial x_j} - \delta_{ij} x_k \frac{\partial^3}{\partial x_k \partial x^2_m}         $$
 
 $$\left[x_i \frac{\partial^2}{\partial x^2_m},\frac{\partial}{\partial x_j} \right]  = -\delta_{ij} \frac{\partial^2}{\partial x^2_m}      $$

 $$\left[x_m \frac{\partial^2}{\partial x_m \partial x_i},\frac{x_j}{r} \right]  = -\left(\delta_{ij} \frac{1}{r} - \frac{x_i x_j }{r^3} + \left( \frac{x_i x_j x_k}{r^3}  - \delta_{ij} \frac{x_k}{r}\right) \frac{\partial}{\partial x_k}\right)         $$
 
 $$\left[x_m \frac{\partial^2}{\partial x_m \partial x_i}, 
 x_j \frac{\partial^2}{\partial x^2_k} \right]= \delta_{ij} x_k \frac{\partial^3}{\partial x_k \partial x^2_m}   -  x_j \frac{\partial^3}{\partial x^2_k \partial x_i}               $$

 $$\left[x_m \frac{\partial^2}{\partial x_m \partial x_i}, x_k \frac{\partial^2}{\partial x_k \partial x_i}\right]  =0   $$
 
 $$\left[x_m \frac{\partial^2}{\partial x_m \partial x_i}, \frac{\partial}{\partial x_j} \right] 
 = - \frac{\partial^2}{\partial x_i \partial x_j}    $$
 
 $$\left[\frac{\partial}{\partial x_i}, \frac{x_j}{r} \right] = \delta_{ij} \frac{1}{r} - \frac{x_i x_j }{r^3}                  $$

$$\left[\frac{\partial}{\partial x_i}, x_j \frac{\partial^2}{\partial x^2_k} \right] = \delta_{ij} \frac{\partial^2}{\partial x^2_m}            $$

$$\left[\frac{\partial}{\partial x_i}, x_k \frac{\partial^2}{\partial x_k \partial x_j}\right] = \frac{\partial^2}{\partial x_i \partial x_j }      
  $$

Substitute the values of the commutator the the original expression. Then

$$ \left[\hat A_i,\hat A_j\right] = F + G  $$

$$  F =  \frac{kh^2}{m} \left( \left[\frac{x_i}{r}, x_j \frac{\partial^2}{\partial x^2_k} \right] - 
\left[\frac{x_i}{r}, x_k \frac{\partial^2}{\partial x_k \partial x_j}\right] - 
\left[\frac{x_i}{r}, \frac{\partial}{\partial x_j}\right] + 
\left[x_i \frac{\partial^2}{\partial x^2_m}, \frac{x_j}{r} \right] -\left[x_m \frac{\partial^2}{\partial x_m \partial x_i},\frac{x_j}{r} \right] -
\left[\frac{\partial}{\partial x_i}, \frac{x_j}{r} \right]  \right) $$    $$=   \frac{kh^2}{m} \left( 2\left( \frac{x_i x_j }{r^3} + \frac{x_i x_j x_k }{r^3}\frac{\partial}{\partial x_k} - \frac{x_j}{r}\frac{\partial}{\partial x_i}\right) - \left( \delta_{ij} \frac{1}{r} - \frac{x_i x_j }{r^3} +\left( \frac{x_i x_j x_k }{r^3} - \delta_{ij} \frac{x_k}{r}\right) \frac{\partial}{\partial x_k}\right) -\right. $$

$$ - \left(-\delta_{ij} \frac{1}{r} + \frac{x_i x_j }{r^3}\right) -2\left( \frac{x_i x_j }{r^3} +  \frac{x_i x_j x_k }{r^3}\frac{\partial}{\partial x_k} -\frac{x_i}{r} \frac{\partial}{\partial x_j} \right) - \left( -\delta_{ij} \frac{1}{r} + \frac{x_i x_j }{r^3} -\left( \frac{x_i x_j x_k }{r^3} - \delta_{ij} \frac{x_k}{r}\right) \frac{\partial}{\partial x_k}\right) -$$ $$ \left.- \left(\delta_{ij} \frac{1}{r} - \frac{x_i x_j }{r^3}\right)\right) =  2k\frac{\hbar^2}{m}\left( \frac{x_i}{r} \frac{\partial}{\partial x_j}  - \frac{x_j}{r}\frac{\partial}{\partial x_i}\right) $$

$$  G =     \frac{\hbar^4}{m^2} \left( \left[x_i \frac{\partial^2}{\partial x^2_m},x_j \frac{\partial^2}{\partial x^2_k}  \right]-
\left[x_i \frac{\partial^2}{\partial x^2_m}, x_k \frac{\partial^2}{\partial x_k \partial x_j} \right]-
\left[x_i \frac{\partial^2}{\partial x^2_m},\frac{\partial}{\partial x_j} \right]-
\left[x_m \frac{\partial^2}{\partial x_m \partial x_i}, 
 x_j \frac{\partial^2}{\partial x^2_k}\right] \right.$$  $$\left.   +
\left[x_m \frac{\partial^2}{\partial x_m \partial x_i}, x_k \frac{\partial^2}{\partial x_k \partial x_j}\right] +
\left[x_m \frac{\partial^2}{\partial x_m \partial x_i}, \frac{\partial}{\partial x_j} \right] 
  - 
\left[\frac{\partial}{\partial x_i}, x_j \frac{\partial^2}{\partial x^2_k} \right] +
\left[\frac{\partial}{\partial x_i}, x_k \frac{\partial^2}{\partial x_k \partial x_j}\right]\right) =  $$  $$ =\frac{\hbar^4}{m^2}\left( 2x_i \frac{\partial^3}{\partial^2 x_k \partial x_j} - 2x_j \frac{\partial^3}{\partial^2 x_k \partial x_i} -  \left(x_i \frac{\partial^3}{\partial^2 x_k \partial x_j} - \delta_{ij} x_k \frac{\partial^3}{\partial x_k \partial x^2_m} \right) +\delta_{ij} \frac{\partial^2}{\partial x^2_m}  -\right. $$ $$ \left.  -\left( \delta_{ij} x_k \frac{\partial^3}{\partial x_k \partial x^2_m}   -  x_j \frac{\partial^3}{\partial x^2_k \partial x_i}\right)  - \frac{\partial^2}{\partial x_i \partial x_j}  -  \delta_{ij} \frac{\partial^2}{\partial x^2_m}+\frac{\partial^2}{\partial x_i \partial x_j }    \right) = \frac{\hbar^4}{m^2} \left(x_i \frac{\partial^3}{\partial^2 x_k \partial x_j} - x_j \frac{\partial^3}{\partial^2 x_k \partial x_i} \right) $$

Hence 
$$\left[\hat A_i,\hat A_j\right] = F +G =  \frac{\hbar^4}{m^2} \left(x_i \frac{\partial^3}{\partial^2 x_k \partial x_j} - x_j \frac{\partial^3}{\partial^2 x_k \partial x_i} \right)+ 2\frac{\hbar^2}{m}\left( \frac{x_i}{r} \frac{\partial}{\partial x_j}  - \frac{x_j}{r}\frac{\partial}{\partial x_i}\right) =  -2i\frac{\hbar}{m}\hat{\hbar}\varepsilon_{ijk} \hat L_k$$

\section{Computation of the $\hat{A}^2$}
Substituting $\hat A_i$  explicitly, we obtain that on a function $\Phi$ as


\begin{equation} \label{sch22}
   \begin{array}{cc}
    &\hat{A}^2 \Phi = \left(\frac{2h^2}{m}\left(\frac{\partial}{\partial x_i} + x_k\frac{\partial^2}{\partial x_k\partial x_i} - x_i\frac{\partial^2}{\partial x^2_n}\right) - k\frac{x_i}{r}\right)
    \left(\frac{2h^2}{m}\left(\frac{\partial}{\partial x_i}
    + x_{\alpha}\frac{\partial^2}{\partial x_{\alpha}\partial x_i} - x_i\frac{\partial^2}{\partial x^2_{\beta}}\right) - k\frac{x_i}{r}\right)\Phi = \\
    & = \left(\frac{2h^2}{m}\right)^2\left(\frac{\partial^2}{\partial x^2_n} + \frac{\partial}{\partial x_i}\left(x_{\alpha} \frac{\partial^2}{\partial x_{\alpha}\partial x_i}\right) - \frac{\partial}{\partial x_i}\left(x_i\frac{\partial^2}{\partial x^2_{\beta}}\right)\right) \Phi - 2\frac{kh^2}{m} \frac{\partial}{\partial x_i}\left(\frac{x_i \Phi}{r}\right) +\\
    & +\left(\frac{2h^2}{m}\right)^2\left(x_k\frac{\partial^3}{\partial x_k \partial x^2_i} + x_k \frac{\partial^2}{\partial x_k \partial x_i} x_{\alpha} \frac{\partial^2}{\partial x_{\alpha} \partial x_i} - x_k \frac{\partial^2}{\partial x_k \partial x_i} x_i \frac{\partial^2}{\partial x^2_{\beta}}\right) \Phi - 2\frac{k \hbar^2}{m} x_k \frac{\partial^2}{\partial x_k \partial x_i} \left(\frac{x_i \Phi}{r}\right) - \\
    &\left(\frac{2h^2}{m}\right)^2 \left(  x_k\frac{\partial^3}{\partial x_k \partial x^2_i} + x_i \frac{\partial^2}{\partial x^2_{\beta}} x_k \frac{\partial^2}{\partial x_k \partial x_i} - x_i \frac{\partial^2}{\partial x^2_n} x_i \frac{\partial^2}{\partial x^2_n}\right) \Phi - 2\frac{k \hbar^2}{m} x_i \frac{\partial^2}{\partial x^2_n}\left(\frac{x_i \Phi}{r}\right) - \\
    & - \frac{2kh^2}{m} \left(\frac{x_i}{r} \frac{\partial}{\partial x_i} + \frac{x_i x_{\alpha} }{r} \frac{\partial^2}{\partial x_i \partial x_{alpha} } - r \frac{\partial^2}{\partial x^2_{\beta}} \right)\Phi +k^2\Phi
\end{array}
\end{equation}

\begin{equation}
    \frac{\partial}{\partial x_i}\left(\frac{x_i \Phi}{r}\right) = \left(\frac{2}{r} + \frac{x_i}{r} \frac{\partial}{\partial x_i} \right) \Phi
\end{equation}

\begin{equation}
     \frac{\partial^2}{\partial x^2_n} \left(\frac{x_i \Phi}{r}\right) = \left( -\frac{2}{r} + r \frac{\partial^2}{\partial x^2_n } \right) \Phi
\end{equation}

\begin{equation}
     \frac{\partial^2}{\partial x_k \partial x_i} \left(\frac{x_i \Phi}{r}\right) = \left( -\frac{2}{r} + \frac{x_i x_k }{r } \frac{\partial^2}{\partial x_i \partial x_k} \right) \Phi
\end{equation}

Hence

\begin{equation}
    \begin{array}{c}
      \hat A^2 = -\left(\frac{\hbar^2}{m}\right)^2 \left( \frac{\partial^2}{\partial x^2_i} + 2 x_i \frac{\partial^3}{\partial x_i \partial x^2_{\beta}} + x_i x_{\alpha } \frac{\partial^4}{\partial x_i \partial x_{\alpha} \partial x^2_{\beta} } - r^2
       \frac{\partial^4}{\partial x^2_n \partial x^2_{\beta}} \right)\\
       - \frac{kh^2}{m} \left( \frac{2}{r} + \frac{2x_i x_k}{r} \frac{\partial^2}{\partial x_i \partial x_k } - 2r \frac{\partial^2}{\partial x^2_n} + 4 \frac{x_i}{r} \frac{\partial}{\partial x_i} \right) +k^2 
     \end{array}
\end{equation}
Because
 \begin{equation}
     \hat{p^2}\hat{L^2} = \hat{L^2} \hat{p^2}
 \end{equation}
we obtain
   \begin{equation}
   \begin{array}{cc}
       \hat A^2 -k^2 &= - \left(\frac{\hbar^2}{m} \right) \frac{\partial^2}{\partial x^2_m} \left( 1 + 2 x_m \frac{\partial}{\partial x_m} - r^2 \frac{\partial^2}{\partial x^2_n} + x_n x_m \frac{\partial^2}{\partial x_n \partial x_m} \right) - 
       \frac{kh^2}{m} \frac{2}{r} \left( 1 + x_i x_k \frac{\partial^2}{\partial x_i \partial x_k } - r^2 \frac{\partial^2}{\partial x^2_n} +2x_i \frac{\partial}{\partial x_i} \right)  =  \\
       &= \frac{1}{m^2} \hat{p^2} \left(\hat{L^2} + \hbar^2\right) - \frac{2}{m} \frac{k}{r} \left(\hat{L^2} + \hbar^2\right) = \frac{2}{m} \left( \frac{\hat{p^2}}{2m} - \frac{k}{r} \right) \left(\hat{L^2} + \hbar^2\right)
       = \frac{2}{m} \hbar \left(\hat{L^2} + \hbar^2\right).
  \end{array}
   \end{equation}


\end{document}